\documentclass[12pt,preprint]{aastex}
\usepackage{natbib}

\begin{document}

\title{On the Appearance of Thresholds in the Dynamical Model of Star Formation}

\author{Bruce G. Elmegreen}
\affil{IBM T. J. Watson Research Center, 1101 Kitchawan Road, Yorktown Heights, New
York 10598 USA} \email{bge@us.ibm.com}

\begin{abstract}
The Kennicutt-Schmidt (KS) relationship between the surface density of the star
formation rate (SFR) and the gas surface density has three distinct power laws that may
result from one model in which gas collapses at a fixed fraction of the dynamical rate.
The power law slope is 1 when the observed gas has a characteristic density for
detection, 1.5 for total gas when the thickness is about constant as in the main disks
of galaxies, and 2 for total gas when the thickness is regulated by self-gravity and
the velocity dispersion is about constant, as in the outer parts of spirals, dwarf
irregulars, and giant molecular clouds. The observed scaling of the star formation
efficiency (SFR per unit CO) with the dense gas fraction (HCN/CO) is derived from the
KS relationship when one tracer (HCN) is on the linear part and the other (CO) is on
the 1.5 part. Observations of a threshold density or column density with a constant SFR
per unit gas mass above the threshold are proposed to be selection effects, as are
observations of star formation in only the dense parts of clouds.  The model allows a
derivation of all three KS relations using the probability distribution function of
density with no thresholds for star formation. Failed galaxies and systems with sub-KS
SFRs are predicted to have gas that is dominated by an equilibrium warm phase where the
thermal Jeans length exceeds the Toomre length. A squared relation is predicted for
molecular gas-dominated young galaxies.
\end{abstract}

\section{Introduction}
\label{introduction}

A correlation between surface density of star formation, $\Sigma_{\rm SFR}$, and
surface density of gas, $\Sigma_{\rm gas}$, is observed in local galaxies on scales
larger than several hundred parsecs \citep[e.g.,][]{kennicutt07} and it is observed for
whole galaxies over a wide range of redshifts (\citealp{daddi10,genzel10}, see review
in \citealp{kennicutt12}). This correlation is typically a power law in the main parts
of spiral galaxy disks with a fall-off in the outer parts \citep{kennicutt89} that is
attributed to a decreasing relative abundance of cool and molecular gas
\citep[e.g.,][]{krumholz13} with star formation following the molecules
\citep{schruba11}. The value of the power law slope is consistently around unity for CO
in normal galaxies \citep{wong02,bigiel08,leroy08} and for dense gas tracers like HCN
\citep{gao04,wu05} and around 1.4 for total gas in the main parts of galaxy disks
\citep{kennicutt98}. It is steeper in the outer regions of spiral galaxies
\citep{bigiel10} and in dwarf irregular galaxies \citep{roy09,bolatto11,eh15} with a
slope of around 2.

The origin of these correlations has been addressed analytically in many previous
studies \citep[e.g.,][]{krumholz05,krumholz09,krumholz12,krumholz13,
ostriker10,hennebelle11,padoan11,renaud12,padoan14} and also shown to follow from
numerical simulations
\citep[e.g.,][]{li05,padoan12,fed12,kim13,kim15,hu16,murante15,semenov16,hopkins17}.
The basic ingredients are gaseous self-gravity, turbulence and cooling, with cloud
geometry and feedback determining the gas scale height and limiting the fraction of gas
that gets into stars.

A recent development is the inference that star formation not only occurs in dense gas,
as long believed, but that the rate of star formation per unit gas mass is independent
of the volume density of this gas \citep[e.g.,][]{evans14}. This inference follows from
the linear correlation between $\Sigma_{\rm SFR}$ and the surface density of dense gas
tracers like HCN \citep{gao04,wu05}, and from a linear correlation between the total
star formation rate (SFR) in molecular clouds and the mass of gas above some extinction
threshold, typically $\sim8$ visual magnitudes \citep[e.g.,][]{lada10}. This
independence from volume density is contrary to expectations based on gravitational
processes, which would imply a local collapse rate proportional to the square root of
the local density on all scales where the gas motions are supersonic. The purpose of
this paper is to question this inference about dense gas and to present an alternative
interpretation in which the apparent threshold is the result of one or more selection
effects.

The existence of a threshold for any aspect of star formation should be questioned in
general. Conceivably, star formation could result from a continuous collapse of
interstellar gas, mitigated somewhat by magnetic fields and stellar feedback, all the
way from the low density atomic medium \citep[e.g.,][]{michalowski15}, which is
unstable on scales less than the Toomre length (Sect. \ref{discussion}), down to the
dense molecular medium and into stars. This is contrary to most theories and
simulations that assume a threshold density either in an integral over the probability
distribution function (PDF) of gas to derive the SFR from first principles
\citep{elmegreen02,kravtsov03,krumholz05} or from numerical constraints which require a
threshold density to convert gas particles into star particles
\citep[e.g.,][]{schaye15,vogelsberger14}. We suggest here that all of the star
formation correlations observed in galaxies could follow from a single model of
pervasive collapse with no thresholds. We show this in two ways, first with the common
dynamical model in which the star formation rate depends on the gas mass divided by the
free fall time, and then using integrals over PDFs for star formation rate, total gas,
molecular gas, and dynamical time, the latter being used as an independent way to get
the molecular gas.

There are several implications of this proposal. First, molecular hydrogen cannot be a
pre-requisite for star formation, but a result of the high densities that happen anyway
during gas collapse \citep[see also][]{glover12,krumholz12b}. This implication changes
the basic equation for star formation \citep[e.g.,][]{krumholz12} by removing a term
proportional to the molecular fraction: collapse begins in the atomic gas at a
large-scale rate that depends primarily the average midplane density, molecular or not.
Second, much of the ISM should have the density structure expected for collapse, namely
a high-density tail of the density PDF that is a power law. Such power laws are
observed in giant molecular clouds \citep[GMCs, e.g.,][see other references in Section
\ref{integrals}]{kainulainen11}, where collapse might be expected anyway, but also in
the whole-galaxy PDF for molecules in M33 \citep{druard14}.  Such pervasive power law
structure is the main expectation of the present model and more observations on a
galactic scale would be interesting. Third, the gas consumption time varies
considerably from place to place, depending on the average gas density and therefore
position in a galaxy or type of galaxy, and is not constant for gas above a threshold
density or column density as currently inferred from dense gas observations. Dense gas
can be stable in high pressure environments and unstable in low pressure environments
(Sect. \ref{integrals}).

Threshold densities for star formation have also been questioned elsewhere.
\cite{gutermuth11} found no thresholds for star formation in observations of the MonR2
and Ophiuchus clouds.  \cite{lada13} and \cite{burkert13} showed that a threshold for
SFR inside molecular clouds may appear because of the combination of a uniformly
increasing SFR per unit area with gas surface density, i.e., a Schmidt relation with a
slope of $\sim2$ (Sect. \ref{ks4}), and a cloud area that varies inversely with surface
density. \cite{parmentier17} explained the apparently constant star formation rate per
unit dense gas mass inside molecular clouds by showing that the total rate integrated
over a cloud with an isothermal density profile is proportional to the dense mass in
the core of the cloud, even though a significant fraction of the stars form outside the
core without an actual density threshold. She also found, for a fixed cloud radius,
that the SFR transitions from a linear dense mass proportionality when the density at
the edge of the cloud is less than the threshold for molecule detection, to a 1.5 power
of dense gas mass at a higher mass because of the extra square root dependence on
density for the collapse rate. \cite{hopkins17} show that the value of a density
threshold used for numerical convenience in a simulation does not affect the star
formation rate on large scales as long as the self-gravitating regions are resolved.
The present study agrees with the conclusions of these others but differs in detail by
deriving the observed KS relations and dense-gas correlations in two ways from a purely
dynamical model of pervasive interstellar collapse.

We begin by summarizing the three distinct Kennicutt-Schmidt relations in Section
\ref{multiple} and then we derive them using the dynamical model without thresholds or
pre-requisites for molecule formation. A second derivation of the three KS laws in
Section \ref{pdfsection} uses integrals over the density PDFs, again without thresholds
for star formation. Unlike other studies, we use a convolution PDF that includes a
power law part at high density. The impact of a column density threshold for the
detection of molecules is in Section \ref{columnthreshold}, and an explanation for the
appearance of threshold densities and column densities for star formation is in Section
\ref{threshold}. A related explanation for the constant star formation rate per unit
dense gas mass is in Section \ref{constant}.  Following a discussion (Sect.
\ref{discussion}) of how this dynamical model fits into a large-scale picture of
interstellar evolution and star formation, the conclusions are in Section
\ref{conclusions}.

\section{Multiple KS Relationships}
\label{multiple}

\subsection{Three Power Laws}

Most studies of the relationship between $\Sigma_{\rm SFR}$ and the gas surface density
have concentrated on the main disks of spiral galaxies, where a difference in slope is
observed for total gas and molecular gas (\citealt{kennicutt98}, \citealt{bigiel08},
see review in \citealt{kennicutt12}).  In the outer parts of spirals and in dIrrs, a
third and steeper slope appears. The transition to this steeper slope occurs where two
things happen simultaneously: the average gas column density becomes less than the
value needed to shield a molecular cloud (e.g., $\sim10\;M_\odot$ pc$^{-2}$; see
Krumholz et al. 2009), and the gas mass begins to dominate the stellar mass, leading to
a flare in the total gas thickness as the surface density decreases with an
approximately constant velocity dispersion \citep{olling96,levine06}. These two things
leave an ambiguity in what drives the steeper slope: is it the sudden lack of molecules
\citep{krumholz13} or the sudden drop of midplane density in the flare
\citep{barnes12,elmegreen15}?

Here we take the viewpoint that a single dynamical law produces the three observed KS
relationships regardless of the pre-existence of molecules with the difference between
them the result of primarily two things: the degree of self-gravity in the gas, which
determines the relation between the density and the column density, and the ratio of
the average density to the characteristic density for radiation of the tracer used to
observe it.

To distinguish between these three KS relationships, we group them according to their
power laws:

\noindent (KS-1a) $\Sigma_{\rm SFR}$ on galactic and sub-galactic scales increases
approximately linearly with the surface density of molecular tracers, such as CO in
normal spirals at low density or HCN at high density
\citep{wong02,gao04,wu05,bigiel08,leroy08,heiderman10,bigiel11,wang11,schruba11,bolatto11,leroy13,
zhang14,liu15,chen17};

\noindent (KS-1b) $\Sigma_{\rm SFR}$ for local molecular clouds scales approximately
linearly with the dense gas surface density determined from extinction or FIR emission
\citep{vuti14};

\noindent (KS-1c) The total star formation rate in a molecular cloud scales about
linearly with the mass of dense gas
\citep{wu10,lada10,lada12,evans14,vuti16,shimajiri17};

\noindent (KS-1.5) the star formation rate surface density, $\Sigma_{\rm SFR}$, scales
with the {\it total} gas surface density, atomic plus molecular, to a power of
approximately $1.5$
\citep[e.g.,][]{buat89,kennicutt89,tenjes91,kennicutt98,liu15,calzetti17};

\noindent (KS-2a) $\Sigma_{\rm SFR}$ scales approximately with the square of the total
gas surface density in the outer regions of spiral galaxies and in dwarf irregular
galaxies \citep[][hereafter Paper I]{roy09,bigiel10,bolatto11,eh15,elmegreen15}.
\citealp{teich16} also found a slope of 2 for low-mass galaxies, but only on large
scales where stochastic variations were smallest;

\noindent (KS-2b) $\Sigma_{\rm SFR}$ scales with the gas surface density in individual
molecular clouds (not just the dense gas) to a power that is approximately $2$
\citep{heiderman10,gutermuth11,harvey13,lada13,evans14,willis15,nguyen16,retes17,lada17}.

There is an offset from the KS-1.5 relationship for some ULIRGs
\citep{greve05,daddi10,genzel10,garcia12}, but that is not viewed here as physically
distinct from KS-1.5 but as a manifestation of either higher densities from
galactic-scale shocks \citep{juneau09,renaud14,kepley16}, a changing molecular
conversion factor \citep{nara12}, or large corrections to the apparent gas mass
\citep{scoville16}.

These relationships follow from the same physical law if we consider that gas can be
observed with different tracers and that it can have different relationships between
the line of sight depth and the surface density. This law is the commonly assumed
dynamical model, which is three-dimensional and written in terms of density, $\rho$, as
\begin{equation}\rho_{\rm SFR}=\epsilon_{\rm ff}\rho/t_{\rm ff},\label{sf3d}\end{equation}
where
\begin{equation} t_{\rm ff}=(32G\rho/[3\pi])^{-1/2}\label{tff}\end{equation}
is the free fall time and $\epsilon_{\rm ff}$ is an approximately constant efficiency
per unit free fall time
\citep[e.g.,][]{larson69,madore77,elmegreen91,elmegreen02,krumholz05}.  In the present
model, there is no term like the molecular fraction, $f_{\rm H2}$, in \cite{krumholz12}
because molecules, molecular clouds, HCN regions etc., are all viewed as incidental and
not causal, so they can be ignored in the equation for star formation, which is
primarily a dynamical process. Observations of $\epsilon_{\rm ff}$ for individual
star-forming regions suggest a range of values consistent with time-variability
\citep{lee16}; numerical simulations get a range for $\epsilon_{\rm ff}$ too
\citep{semenov16}.

A summary of analytical models that reproduce these relationships follows. Some are
newly derived and all follow from equation (\ref{sf3d}) without thresholds.

\subsection{KS-1.5: Star formation for total gas in main spiral galaxy disks}

KS-1.5 is the standard relation for star formation in galaxies \citep{buat89,
kennicutt89,kennicutt98} so we begin with that here. It follows from equation
(\ref{sf3d}) if the observed disk region has an approximately constant scale height,
$H$, as observed for CO in the Milky Way \citep{heyer15}.  Then $\rho=\Sigma/2H$ and
\begin{equation}\Sigma_{\rm SFR}=\epsilon_{\rm ff}(16G/[3\pi H])^{1/2}\Sigma_{\rm
gas}^{3/2}.\label{kstotal}
\end{equation}
With typical $H=100$ pc and $\epsilon_{\rm ff}=0.01$, this becomes
\begin{equation} {{\Sigma_{\rm SFR}}\over{M_\odot\;{\rm pc}^{-2}\;{\rm
Myr}^{-1}}}= 8.8\times10^{-5}\left({{\Sigma_{\rm gas}}\over{M_\odot\;{\rm
pc}^{-2}}}\right)^{1.5}. \label{eq:total0}
\end{equation}
This result was shown in Paper I to agree in both slope and intercept with the
observations in \cite{kennicutt12}, which are for average star formation rates in the
main disks of spiral galaxies \citep[the average rate per unit area in a galaxy is a
good reflection of the local rate;][]{elmegreen07a}.

The approximately constant $H$ requires a separate model with a more complete theory of
interstellar processes \cite[e.g.,][]{kim15}. For example, the squared velocity
dispersion of the gas, $\sigma^2$, should be proportional to the total mass surface
density in the gas layer because $H=\sigma^2/(\pi G\Sigma_{\rm tot})$ is then constant.
Such a relation may be a consequence of the KS-1.5 relation if we consider that the
energy density decay rate per unit area in the gas, $0.5\Sigma_{\rm gas}\sigma^3/H$
\citep[for dissipation in a crossing time $H/\sigma$;][]{maclow98,stone98}, is
proportional to the product of the gas surface density and the SFR, $\propto\Sigma_{\rm
gas}\Sigma_{\rm SFR}$. Then $\Sigma_{\rm gas}\sigma^3/H=\pi G\Sigma_{\rm
gas}\Sigma_{\rm tot}\sigma \propto\Sigma_{\rm gas}\Sigma_{\rm SFR} \propto\Sigma_{\rm
gas}^{2.5}$ so that $\sigma\propto\Sigma_{\rm gas}^{0.5}\left(\Sigma_{\rm
gas}/\Sigma_{\rm tot}\right)$ and $H\sim\left(\Sigma_{\rm gas}/\Sigma_{\rm
tot}\right)^3$. This ratio for $H$ is about constant in the main disks of spiral
galaxies where both stars and gas have the same exponential scale lengths.

\subsection{KS-1a: Star formation in molecular gas at the characteristic density for emission}
\label{sfrinhcn}

\subsubsection{Derivation from Analytical Theory}

KS-1a follows for a constant effective density for emission, $\rho_{\rm mol}$
\citep{evans99,shirley15,jimenez17,leroy17a}, that is much larger than the average
interstellar density, $\rho_{\rm gas}$. Then equation (\ref{sf3d}) converts to
\begin{equation}
\Sigma_{\rm SFR}=\epsilon_{\rm ff}\Sigma_{\rm mol}/t_{\rm ff,mol}
\label{ksmol}
\end{equation}
for constant $t_{\rm ff,mol}$ equal to the free fall time at the characteristic density
for emission by the molecule (which could be CO, HCN, or some other tracer of a
particular phase of gas). Here we have set the fraction of the interstellar medium in
the molecular phase equal to the fraction of the time spent as molecules, based on the
local dynamical time (Paper I).

\begin{equation}
f_{\rm mol}={{\rho_{\rm mol}^{-0.5}}\over{\rho_{\rm mol}^{-0.5}+\rho_{\rm gas}^{-0.5}}}\sim
\left({{\rho_{\rm gas}}\over{\rho_{\rm mol}}}\right)^{0.5}
\label{eq:timing}
\end{equation}
(the approximation is for partially molecular regions, $\rho_{\rm gas}<<\rho_{\rm
mol}$). This expression assumes that both the cloud formation time before star
formation and the cloud break-up time after star formation are proportional to the
local dynamical times, in agreement with numerical simulations of a supernova-agitated
interstellar medium \citep{padoan16a}.  Then
\begin{equation}\Sigma_{\rm mol}=f_{\rm mol}\Sigma_{\rm gas}\end{equation}
and equation (\ref{ksmol}) follows from equation (\ref{sf3d}).

\subsubsection{KS-1a in Comparison to Observations}

To evaluate equation (\ref{ksmol}) in case KS-1a, we consider that CO appears in local
clouds at about 1.5 magnitude of visual extinction \citep{pineda08}, which corresponds
to $\sim30\;M_\odot$ pc$^{-2}$ of column density. For a typical large cloud near this
extinction threshold with a size of $\sim30$ pc, the 3D density is $\sim17$ H$_2$
cm$^{-3}$ and for this density $t_{\rm ff,mol}=8.0$ Myr. With $\epsilon=0.01$ again,
equation (\ref{ksmol}) becomes
\begin{equation} {{\Sigma_{\rm SFR}}\over{M_\odot\;{\rm pc}^{-2}\;{\rm
Myr}^{-1}}}= 1.2\times10^{-3}\left({{\Sigma_{\rm mol,CO}}\over{M_\odot\;{\rm
pc}^{-2}}}\right) \label{eq:total1}
\end{equation}
giving a consumption time of $0.80$ Gyr. This molecular consumption time is too short
by a factor of $\sim2$ compared to star formation rates on a large scale
\citep[e.g.,][]{bigiel08,leroy08}, suggesting that $\epsilon_{\rm ff}/t_{\rm ff}$
should be lower by this factor. For example, \cite{leroy17b} find $\epsilon_{\rm
ff}\sim0.003$ on 40 pc scales in M51 and discuss how $\epsilon_{\rm ff}$ is often
observed to be lower than 0.01. \cite{murray11} suggest $\epsilon_{\rm ff}\sim0.006$ on
average in the Milky Way.

\cite{leroy17b} considered star formation relationships for CO observations at 40 pc
resolution in M51. They found that the molecular depletion time, which is $\Sigma_{\rm
mol}/\Sigma_{\rm SFR}=t_{\rm ff,mol}/\epsilon_{\rm ff}$ in our notation, is
approximately constant instead of their expected $\Sigma_{\rm mol}^{-0.5}$ for
dynamical star formation at fixed $\epsilon_{\rm ff}$. However, this depletion time
should be constant if the average density for observations of CO is constant, as above,
because that gives the linear molecular relation, KS-1a.  The stated expectation was
that the density used for the dynamical time would be proportional to the average
density in the 40 pc region, but that is not the case if the average CO density in the
resolution element is less than the characteristic density for CO emission.  The
average density comes from the summed mass of the CO clouds in the 40 pc region, but
each cloud could have about the same characteristic density for CO emission and the
same $t_{\rm ff,mol}$. The best correlation they found was with the virial parameter,
$5R\sigma^2/(GM)$ for $R=40$ pc and mass $M$ inside the region. They determined that
the depletion time scales with the virial parameter to a power of $\sim0.9$. At the
same time, \cite{leroy17b} found that $\epsilon_{\rm ff}$ is nearly independent of the
virial parameter. These two results imply, for the dynamical model, that the average
density per molecular cloud, which occurs inside $t_{\rm ff,mol}$, depends on the
average virial parameter measured on the scale of 40 pc.

In the case of a dense molecular tracer, like HCN or HCO$^+$, the characteristic
density of observation is $\sim3\times10^4$ cm$^{-3}$ in equation (\ref{ksmol}), giving
$t_{\rm ff,mol}=0.19$ Myr and with $\epsilon=0.01$,
\begin{equation} {{\Sigma_{\rm SFR}}\over{M_\odot\;{\rm pc}^{-2}\;{\rm
Myr}^{-1}}}= 0.052\left({{\Sigma_{\rm mol,HCN}}\over{M_\odot\;{\rm
pc}^{-2}}}\right) \label{eq:total2}
\end{equation}
The average observed coefficient is slightly lower than 0.052, i.e., more like 0.02
(Sect. \ref{constant}), so $\epsilon_{\rm ff}$ is proportionally lower or the
characteristic density for emission is slightly higher. The result is close enough to
the observation to support the general model, given the uncertainties in density, star
formation rate, and dense mass, plus the approximate nature of the model itself.

At high interstellar density (which usually corresponds to high $\Sigma_{\rm gas}$),
$\rho_{\rm gas}\gtrsim\rho_{\rm mol}$ and $f_{\rm mol}\sim1$, in which case equation
(\ref{kstotal}) applies with $\Sigma_{\rm mol}\sim\Sigma_{\rm gas}$. Thus molecular
emission has a 1.5 power law at high $\rho_{\rm gas}$ (if $H$ is still about constant)
and a linear law at interstellar densities below the effective density for emission
where $f_{\rm mol}<1$ \citep[e.g.,][]{kennicutt98,krumholz12}. For example,
\cite{gowardhan17} got a slope of $1.41\pm0.10$ for CO emission in ULIRGS and high
redshift galaxies where the density is large ($\rho_{\rm gas}\gtrsim\rho_{\rm mol}$ for
CO), and they got a slope closer to unity, $1.11\pm0.05$, for the dense gas tracer HCN
in the same galaxies, presumably because $\rho_{\rm gas}\lesssim\rho_{\rm mol}$ and
$f_{\rm mol}<1$ for HCN in these galaxies. \citep[For more discussion on this point,
see][]{krumholz07,nara08,burkert13,elmegreen15}.

\subsubsection{The Relationship between Star Formation Efficiency and Dense Gas Fraction}
\label{section:sfe}

An important correlation appears for average interstellar densities that are between
the characteristic densities for CO and HCN observations (or any other low and high
density tracers). Above the CO density, CO tracks the total interstellar density fairly
well because $f_{\rm mol,CO}\sim1$ and then $\Sigma_{\rm SFR}\propto\Sigma_{\rm
CO}^{1.5}$ for a constant thickness galaxy, as mentioned above. Below the HCN density,
$f_{\rm mol,HCN}<1$ for the average interstellar medium and then the linear law
appears: $\Sigma_{\rm SFR}\propto\Sigma_{\rm HCN}$. Thus the ratio of HCN to CO, which
is viewed as the ``dense gas fraction,'' increases with the star formation rate,
\begin{equation}
f_{\rm dense}={{\Sigma_{\rm HCN}}\over{\Sigma_{\rm CO}}}\propto\Sigma_{\rm SFR}^{1/3}
\propto\Sigma_{\rm CO}^{1/2}.
\end{equation}
Similarly, the ``star formation efficiency'', measured as the ratio $\Sigma_{\rm
SFR}/\Sigma_{\rm CO}$, should scale linearly with $f_{\rm dense}$:
\begin{equation}
{\rm SFE} = {{\Sigma_{\rm SFR}}\over{\Sigma_{\rm CO}}} \approx
{{{\Sigma_{\rm CO}}^{3/2}}\over{\Sigma_{\rm CO}}}=
\Sigma_{\rm CO}^{1/2}\approx\Sigma_{\rm SFR}^{1/3}\approx f_{\rm dense}
\label{sfe}
\end{equation}
These correlations are consistent with observations of star-forming galaxies and ULIRGs
in \cite{usero15} and elsewhere. Similarly, \cite{gao04} observed $L_{\rm HCN}/L_{\rm
CO}\propto L_{\rm CO}^{0.38}$ which is similar to our prediction of a power of
$\sim0.5$ in this middle-density regime at constant $H$, and they observe $L_{\rm
IR}/L_{\rm CO}\propto \left(L_{\rm HCN}/L_{\rm CO}\right)^{1.24}$, which is similar to
our predicted power of 1. Sections \ref{columnthreshold} and \ref{constant} return to
discuss star formation in dense gas, including sublinear relations between SFR and HCN
which are not considered above.

\subsection{KS-2a: Star Formation in Total Gas for dIrrs and Outer Spiral Galaxy Disks}
\label{dirrsf}

KS-2 follows when the line-of-sight thickness of the region is in pressure equilibrium
with gas self-gravity for the observed gas column density. This should be the case in
the gas-dominated parts of galaxy disks. We set the disk scale height $H=\sigma^2/\pi G
\Sigma$ and then derive (Paper I)
\begin{equation}\Sigma_{\rm SFR}=(4/\sqrt{3})\epsilon_{\rm ff}G\Sigma_{\rm
gas}^2/\sigma.\label{eq:dwarf}
\end{equation}
With a constant velocity dispersion $\sigma=6$ km s$^{-1}$ as typically observed in
dwarf irregulars and outer spiral disks, and for $\epsilon_{\rm ff}=0.01$,
\begin{equation}
{{\Sigma_{\rm SFR}}\over{M_\odot\;{\rm pc}^{-2}\;{\rm Myr}^{-1}}}
=1.7\times10^{-5}
\left({{\Sigma_{\rm gas}}\over{M_\odot\;{\rm
pc}^{-2}}}\right)^2 .
\label{eq:outer}
\end{equation}
This relation was shown in Paper I to agree with observations of the outer parts of
spiral disks and dwarf irregular galaxies. The main reason for the steepening of the
slope is the increase in scale height with radius, i.e., the disk flare in a galaxy.
That increase drops the midplane gas density faster than the surface density so the
dynamical rate at the midplane density drops more quickly too. A disk flare was also
present in the \cite{krumholz13} model although not mentioned explicitly.

These regions of low surface brightness are also where the metallicity tends to be low
\citep{rosales12,bresolin15}, but the drop in $\Sigma_{\rm SFR}$ is probably not from
an inability to make H$_2$ on dust. This is because the star formation relation in this
regime is the same for a wide range in metallicities, i.e., comparing outer spiral
disks where the metallicity is slightly below solar to dwarf irregular galaxies, where
the metallicity is $\sim10$\% solar \citep{roy15,jameson16}.  The squared dependence of
$\Sigma_{\rm SFR}$ on $\Sigma_{\rm gas}$ is also not from a drop in molecular fraction
with decreasing density because the density dependence of the molecular fraction for
conventional theory \citep{krumholz09} is much steeper than the observed decrease in
$\Sigma_{\rm SFR}$ with gas density in galaxies \citep{eh15}.

\cite{ostriker11} derived a squared KS relation on a galactic scale by assuming that
the interstellar pressure is proportional to $\Sigma_{\rm SFR}$ through momentum
injected by supernovae, and that this pressure is also proportional to $\Sigma_{\rm
gas}^2$ as in an equilibrium galaxy disk. The application of supernova regulation in
outer spiral disks and dIrr galaxies is not clear though, considering the very low star
formation rate and pressure there. For example, in an exponential disk, the surface
density of supernovae decreases as $\exp(-R/R_{\rm D})$ for star formation scale length
$R_{\rm D}$, and the midplane gas density decreases as $\rho\propto\exp(-2R/R_{\rm D})$
considering the outer-disk flare (i.e., $\rho=(\pi/2)G\Sigma^2/\sigma^2$ for an
equilibrium disk of pure gas with a near-constant velocity dispersion $\sigma$ and the
same exponential for gas surface density, $\Sigma$). Considering that the radius at
which a supernova remnant merges with the ambient medium scales as $\rho^{-3/49}$
\citep{cioffi88}, it follows that the volume filling factor of remnants, which is this
radius cubed multiplied by the space density of supernovae, decreases with
galactocentric radius as
\begin{equation}
f_{\rm SNR}\sim e^{-R/R_{\rm D}}e^{18R/49R_{\rm D}}\sim e^{-0.63R/R_{\rm D}}.
\end{equation}
Thus, outer galaxy disks should have relatively sparse stirring by supernovae. A
flatter decrease than this for the total gas surface density, e.g., from an extended HI
disk, makes this conclusion even stronger.

\subsection{KS-2b: Star Formation on the Molecular Cloud Scale}
\label{ks4}

KS-2b may follow from the same relationship as KS-2a if it is applied to the interiors
of self-gravitating clouds or to whole self-gravitating clouds.  We assume a power-law
internal density profile $\rho(r)=\rho_{\rm edge}(r_{\rm edge}/r)^\alpha$ from some
small core radius, $r_{\rm core}$ to the edge radius $r_{\rm edge}$ where the density
is $\rho_{\rm edge}$ (an internal profile that explicitly includes $\rho_{\rm core}$ is
in equation (\ref{eq:rho})). This gives a radius-dependent mass
\begin{equation}
M(r)={{4\pi}\over{3-\alpha}}\rho_{\rm edge}r_{\rm edge}^\alpha r^{3-\alpha}
\end{equation}
and surface density $\Sigma(r)=M(r)/(\pi r^2)$. As an approximation, we take the
one-dimensional velocity dispersion $\sigma(r)$ from the virial theorem,
\begin{equation}
3\int_0^r \sigma(r)^2 \rho(r)4\pi r^2 dr=\int_0^r (GM[r]/r)\rho(r)4\pi r^2dr,
\end{equation}
which gives
\begin{equation}
\sigma(r)^2={{4\pi G}\over{3(3-\alpha)}}\rho_{\rm edge}r_{\rm edge}^\alpha r^{2-\alpha}=GM(r)/(3r).
\end{equation}
The internal surface density for the SFR then follows from equation (\ref{sf3d}), which
is also a function of radius,
\begin{equation}
\Sigma_{\rm SFR}(r)={{1}\over{\pi r^2}}\int_0^r\left({{\epsilon_{\rm ff}\rho(r)}\over
{t_{\rm ff}(r)}}\right) 4\pi r^2dr.
\end{equation}
For a singular isothermal sphere, $\alpha=2$, the SFR surface density has a logarithmic
divergence near the center of the cloud, which requires the use of a core radius,
\begin{equation}
\Sigma_{\rm SFR}(r)=\sqrt{8/9}\epsilon_{\rm ff}\left({{G\Sigma(r)^2}\over{\sigma(r)}}\right)
\ln\left(r/r_{\rm core}\right).
\label{ksgmc1}
\end{equation}
For other $\alpha<2$,
\begin{equation}
\Sigma_{\rm SFR}(r)=\sqrt{8/9}\left({{(3-\alpha)^{1.5}}\over{3-1.5\alpha}}\right)
\epsilon_{\rm ff}\left({{G\Sigma(r)^2}\over{\sigma(r)}}\right).
\label{ksgmc2}
\end{equation}
We assume for comparison with observations that $\ln(r/r_{\rm
core})\sim(3-\alpha)^{1.5}/(3-1.5\alpha)\sim2.45$, which follows from the first
expression when $\alpha=2$ if $r/r_{\rm core}=11.6$ and from the latter expression if
$\alpha=1.5$, which are two cases considered also in Section \ref{pdfsection}. Then
\begin{equation}
\Sigma_{\rm SFR}(r)\approx2.31\epsilon_{\rm ff}G\Sigma(r)^2/\sigma(r).
\label{ksgmc}
\end{equation}
These results were written to resemble equation (\ref{eq:dwarf}) in form.

Table 2 in \cite{gutermuth11} gave fits for the $\Sigma_{\rm SFR}-\Sigma_{\rm gas}$
relation for 8 local molecular clouds. Averaging their coefficients in the log and
averaging their powers of $\Sigma_{\rm gas}$, we get from their observations
\begin{equation}
{{\Sigma_{\rm SFR}}\over{M_\odot\;{\rm pc}^{-2}\;{\rm Myr}^{-1}}} \approx3.6\times10^{-4}
\left({{\Sigma_{\rm gas}}\over{M_\odot\;{\rm pc}^{-2}}}\right)^{2.2}
\label{guter}
\end{equation}
This observation agrees with equation (\ref{ksgmc}) at $\Sigma_{\rm gas}=100\;M_\odot$
pc$^{-2}$ and $\sigma=1$ km s$^{-1}$ if $\epsilon_{\rm ff}=0.09$. This efficiency
inside molecular clouds is factor of $\sim9$ larger than it is on a galactic scale but
perhaps this excess is reasonable if the galactic scale contains CO gas that is not
connected with SF.

\cite{lada13} fitted the star formation rates in three local clouds, Orion A, Taurus
and California Nebula,
\begin{equation}
{{\Sigma_{\rm SFR}}\over{M_\odot\;{\rm pc}^{-2}\;{\rm Myr}^{-1}}} \approx4.6\times10^{-5}
\left({{\Sigma_{\rm gas}}\over{M_\odot\;{\rm pc}^{-2}}}\right)^{2.0},
\end{equation}
which is a factor of $\sim8$ lower than in equation (\ref{guter}). This would translate
to $\epsilon_{\rm ff}=0.005$ in equation (\ref{ksgmc}) for $\sigma\sim1$ km s$^{-1}$.
These studies determine the local star formation rates from counts of young stars,
which involve assumptions about stellar ages and masses, and there could also be
stochastic effects for low counts. \cite{lada13} note that the relation is steeper for
Orion B, where the slope is 3.3. The real relation should scale with $\Sigma_{\rm
gas}^2/\sigma$, however, and $\sigma$ may vary with $\Sigma_{\rm gas}$ (see below).

\cite{gutermuth11} considered a reason for their molecular cloud relation that is
somewhat like ours, deriving a KS-2b relation based on counting the areal density of
thermal Jeans mass objects in a thin self-gravitating cloud. Equation (\ref{ksgmc}) is
more general and shows directly the connection between the molecular cloud relation and
the galactic relation for gas-dominated regions (KS-2a). \cite{parmentier13} derived
approximately the same squared star formation law for molecular clouds as in equation
(\ref{ksgmc}) using the area integral over a collapse-model density profile \citep[see
also][]{parmentier11}. They applied it to the formation of star clusters.

Other observations of the KS relation on molecular cloud scales are in
\cite{heiderman10}, \cite{harvey13}, \cite{lada13}, \cite{willis15}, \cite{heyer16},
\cite{nguyen16}, \cite{retes17} and \cite{lada17}, who all found a power index of
$\Sigma_{\rm gas}$ close to 2 (or sometimes larger) and a similar factor of $\sim20$
efficiency compared to the galactic-scale KS law.

\cite{evans14} considered young stellar objects in 25 local clouds using the c2d and
Gould Belt {\it Spitzer} legacy program \citep{evans03,dunham13}. They plotted the KS
relation for these clouds with 3D density instead of column density, using the same
thickness for the star formation rate and the gas, so it is essentially the same as
plotting surface densities. They got a slope of $2.02\pm0.07$ and suggested that this
slope is inconsistent with the dynamical star formation model. That inconsistency is
only in comparison to KS-1.5, however, with its slope of 1.5 for main galaxy disks.

Equation (\ref{ksgmc}) should also apply to whole molecular clouds if the radius is
taken to be the cloud radius so the average surface density and cloud dispersion are
used. For the \cite{evans14} data with tabulated values of $\Sigma_{\rm SFR}$,
$\Sigma_{\rm gas}$ and $\sigma$, we derive a proportionality constant $\epsilon_{\rm
ff}=0.016$ for equation (\ref{ksgmc}). Another survey of star-forming complexes was
made for the Large Magellanic Cloud \citep{ochsendorf17}. The average value of
$\Sigma_{\rm SFR}/\left(\Sigma_{\rm gas}^2/\sigma\right)$ for that implies
$\epsilon_{\rm ff}=0.04$. These values are reasonably consistent with other values of
$\epsilon_{\rm ff}$ discussed in this paper, considering the difficulty in defining the
local star formation rate in a molecular cloud.

\cite{heyer16} saw no KS relationship for young stellar objects in Milky Way molecular
clouds because the range in $\Sigma_{\rm gas}$ was too small (their figure 10a), but
they compared it only with the galactic $\sim1.5$ law and the galactic linear molecular
law. The plotted points actually show a steeper relationship on average, consistent
with a slope of 2 or 3.  \cite{heyer16} did find an approximately linear correlation
between $\Sigma_{\rm SFR}$ and $\Sigma_{\rm gas}/t_{\rm ff}$, however, and this is the
basic model assumed in equation (\ref{ksgmc}).

\cite{wu10} compared the total star formation rates in molecular clouds, $dM_{\rm
star}/dt$, with the local rates in the dynamical model, $\rho_{\rm gas}^{1.5}$, and got
a decreasing relationship in their Figure 35 which they claimed was inconsistent with
an expected positive correlation in the dynamical model. However, they should have
compared the total rate with the product of the cloud volume times the local rate,
which would have introduced an additional term $\left(\Sigma_{\rm gas}/\rho_{\rm
gas}\right)^3$ for cloud volume to be multiplied by the local rate $\rho_{\rm
gas}^{1.5}$. The result would have been proportional to $\Sigma_{\rm gas}^3\rho_{\rm
gas}^{-1.5}$ as in their observed decreasing relation, considering that their
$\Sigma_{\rm gas}$ had a narrow range. Thus their result is also consistent with
molecular cloud evolution on a self-gravitating timescale.  Their explanation for the
decreasing relationship is that cloud mass and therefore IR luminosity is correlated
inversely with density, as found by \cite{larson81}, but the KS relationship should be
between IR luminosity per unit volume and gas density (or IR luminosity per unit area
and gas surface density).

The small-scale mechanism of star formation inside molecular clouds is not addressed by
the simple dynamical model of equation (\ref{ksgmc}).  Molecular clouds appear to be
composed of numerous filaments \citep{andre10,molinari10} and it might be that
collisions between these filaments trigger local star formation
\citep{myers09,schneider12}.  \cite{parmentier17} considered the cloud KS law with
filamentary extensions to large radius.  The density PDF for filamentary structure has
been considered by \cite{myers15}.  If mutual gravity is involved, causing the
filaments to collide with each other, then the dynamical model should still apply
because it states only that the rate of star formation on a large scale, i.e., averaged
over many filaments, is proportional to the rate of mutual gravitational attraction.

\subsection{Other Considerations}

\cite{krumholz12} suggested a two-regime model for KS-1.5 using the line-of-sight
integrated form of equation (\ref{sf3d}) where $\Sigma_{\rm SFR}$ is proportional to
$\Sigma_{\rm gas}/t_{\rm ff}$. In one regime, the density used for $t_{\rm ff}$ was the
3D cloud density of a Jeans-mass cloud with a fixed surface density, $\Sigma_{\rm
GMC}$, and in another regime, the density for $t_{\rm ff}$ was the average disk value
when Toomre $Q=1$. The value of $t_{\rm ff}$ used for the relationship was the smaller
of these two. While this method gave a good fit to the data, we consider that clouds
with a fixed $\Sigma_{\rm GMC}$ are more or less star-forming depending on the local
interstellar pressure (see equation \ref{sigmathres} below), and that $Q\sim$constant
is not a dependable criterion for interstellar properties (Sect. \ref{discussion}).
Still, the utility of a KS law written explicitly in terms of $\Sigma_{\rm gas}/t_{\rm
ff}$ rather than $\Sigma_{\rm gas}$ alone, or one written in terms of $\Sigma_{\rm
gas}^2/\sigma$ for self-gravitating regions, is that important dynamical processes can
be considered when $t_{\rm ff}$ or $\sigma$ are included in addition to the total
available gas for star formation.

KS-1.5 and KS-1 break down on small scales because star formation and cloud evolution
are time dependent, and observations on small scales no longer see the average values
that are used in the simple theory reviewed here \citep{schruba10,kruijssen14}.

KS-1.5, KS-1, and KS-2a also contain another, hidden, relation that is independent of
star formation and that is the simultaneous radial decrease of both $\Sigma_{\rm SFR}$
and $\Sigma_{\rm gas}$ from the exponential profile of a galaxy disk \citep{bolatto17}.
This dependence stretches out the relation to cover a large range in both quantities.
Azimuthal variations from spiral arms also contain an independent relation that
stretches out the parameter range. Spiral arms collect and disperse molecular gas and
its associated star formation without much of a change in the star formation rate per
unit molecular mass \citep{ragan16,schinnerer17}.  These additional dependencies can be
important in some cases. For example, azimuthal variations of $\Sigma_{\rm gas}$ in
dwarf irregular galaxies may not have associated variations in the disk thickness and
then KS-1.5 or KS-1a would apply to those variations, while KS-2a still applies in the
radial direction as the thickness increases.

The dense gas relations KS-1b and KS-1c do not follow from the dynamical model where
cloud evolution is always proportional to the gravitational collapse rate. We suspect
KS-1b and KS-1c may be artifacts of observational selection, as discussed in Sections
\ref{columnthreshold}, \ref{threshold} and \ref{constant}.

\section{Power law Probability Distribution Functions and their Role in the KS
Relation} \label{pdfsection}
\subsection{Integrals over the Convolution PDF Function}
\label{integrals}

The previous section showed how observations of star formation in total gas or in CO
and HCN molecules follow from equation (\ref{sf3d}) for main and outer galaxy disks,
dIrrs, and individual GMCs.  Another way to derive the SFR has been to use an integral
over the interstellar density PDF above some threshold density, assuming the PDF is a
log-normal \citep{elmegreen02,kravtsov03,krumholz05}.  This use of a threshold density
assumes that all of the gas above the threshold is involved with the star formation
process and is therefore strongly self-gravitating.  This cannot be the case for a
fixed threshold, however. What matters is the virial parameter, $5R\sigma^2/GM$, for a
particular interstellar cloud of radius $R$, velocity dispersion $\sigma$ and mass $M$
\citep{padoan12}. This parameter has to be less than about unity for strong
self-gravity. That condition is the same as requiring a threshold (minimum) column
density that depends only on the ambient pressure $P$:
\begin{equation}
\Sigma_{\rm Threshold}=\left({{2P}\over{\pi G}}\right)^{0.5}.
\label{sigmathres}
\end{equation}
(given that the pressure in a self-gravitating cloud is $\pi G\Sigma^2/2$ for cloud
column density $\Sigma$).

In high-pressure regions of galaxies, such as the inner regions where both the stellar
and the gas surface densities are high (the pressure scales with the product of these
two quantities), regions with a certain fixed column density like 8 magnitudes of
visual extinction will not be self-gravitating. They will be like the diffuse clouds
observed locally where the pressure is low. Similarly, in the far-outer regions of
galaxies or in dwarf irregulars and low surface-brightness galaxies, where the pressure
is low, clouds with even modest column densities can be self-gravitating if they exceed
the threshold given by equation (\ref{sigmathres}) \citep[see also the discussion in][]
{elmegreen13}. A fixed column density or, correspondingly, a fixed density at a given
cloud mass, cannot be part of a universal condition for star formation. Thus we
consider here the KS relations using density PDFs without threshold densities, and we
use power-law PDFs as recently observed on the cloud scale and possibly larger
\citep[e.g.][]{druard14}.

Turbulence that randomly compresses and decompresses gas takes the local density on a
random walk in the log of the density, and this makes the PDF of density a log-normal
\citep{vaz94,nordlund98}. Interstellar gas that has some density structure in addition
to local turbulence should have a PDF equal to the convolution of that structure with
the log-normal from local turbulence \citep{elmegreen11}. This implies that gravitating
gas with power-law density gradients in dense cores and filaments should have a
power-law PDF, as simulated \citep{klessen00,vaz08,kritsuk11,fed13,pan16} and observed
with dust extinction \citep{froebrich10,kainulainen11}, dust emission
\citep{schneider13,lombardi15,schneider15a,schneider15c} and molecular line emission
\citep{schneider16}. \cite{schneider15b} found a power law characteristic of collapse
up to $A_{\rm V}\sim100$ mag and then a flatter power law beyond, which they supposed
was from some termination of the collapse. \cite{schneider12} determined the PDF of the
Rosette molecular cloud and suggested it had an extension to higher density because of
compression from the nebula.

In a steady-state, collapse-like motions have a local velocity, $v(r)$, proportional to
$(GM(r)/r)^{0.5}$ for mass $M(r)$ inside radius $r$, and these motions produce a
density gradient toward the collapse center, $\rho(r)$, that makes the inflow flux,
$4\pi r^2 v(r) \rho(r)$, approximately constant. The solution to these equations is
$\rho\propto r^{-2}$.  A singular isothermal sphere in virial equilibrium also has this
density profile, while a collapsing envelope onto a core can have a $\rho(r)\propto
r^{-3/2}$ profile \citep{shu77}. Observations of these density gradients in dense cores
mapped by sub-mm wave dust emission were in \cite{mueller02}.

In general, if $\rho\propto r^{-\alpha}$ then the PDF slope is $-3/\alpha$
\citep{kritsuk11, elmegreen11}. This result may be derived for $\rho\propto
r^{-\alpha}$ from the expression where density and radius correspond one-to-one,
$P_{\rm PDF,3D}(\rho)d\rho=P_{\rm 3D}(r)dr$, which gives \citep[e.g., see
also][]{fed13,schneider13,girichidis14}
\begin{equation}
P_{\rm PDF,3D}(\rho)=P_{\rm 3D}(r)/(d\rho/dr).
\end{equation}
Considering that the probability of radius $r$ is $P_{\rm 3D}(r)=4\pi r^2$ and that
$d\rho/dr\propto -r^{-\alpha-1}$, this gives $P_{\rm PDF,3D}\propto
r^{\alpha+3}\propto\rho^{-1-3/\alpha}$ for equal intervals of $\rho$ and
$\rho^{-3/\alpha}$ for equal intervals of $\ln\rho$.

In the same way, the power in the PDF for surface density may be obtained from the
one-to-one relation $P_{\rm PDF,2D}(\Sigma)d\Sigma=P_{\rm 2D}(b)db$ for impact
parameter $b$.  With $\Sigma\propto b^{-\beta}$ and $P_{\rm 2D}\propto b$, the result
is $P_{\rm PDF,2D} \propto b^{\beta+2}\propto\Sigma^{-1-2/\beta}$ for equal intervals
of $\Sigma$ and $\Sigma^{-2/\beta}$ for equal intervals of $\ln\Sigma$.  For a
spherical cloud, $\beta=\alpha-1$, so if $\alpha=2$, then $\beta=1$ and $P_{\rm PDF,2D}
\propto \Sigma^{-2}$ for equal $\ln\Sigma$ intervals.  For $\alpha=3/2$, $P_{\rm
PDF,2D} \propto \Sigma^{-4}$ for equal $\ln\Sigma$ intervals.

In a more general situation \citep{elmegreen11}, the total 3D PDF may be approximated
by the convolution of the average radial density profile, $\rho_{\rm ave}(r)$, and the
local PDF representing fluctuations around this average, $P_{\rm PDF,local}$:
\begin{equation}
P_{\rm PDF,3D}(\rho)=\int_{\rho_{\rm ave\;min}}^{\rho_{\rm ave\;max}}
P_{\rm PDF,local}(\rho | \rho_{\rm ave})P_{\rm ave}(\rho_{\rm ave})
d\rho_{\rm ave}.\label{eq1}
\end{equation}
$P_{\rm PDF,local}(\rho | \rho_{\rm ave})$ is the conditional probability distribution
function for density $\rho$, given the average $\rho_{\rm ave}$. We assume a local PDF
from supersonic turbulence:
\begin{equation}P^{\prime}_{\rm PDF,local}(\rho)=(2\pi D^2)^{-1/2}e^{-0.5\left(\ln(
\rho/\rho_{\rm pk})/D\right)^2},
\end{equation}
where $P^{\prime}$ indicates the function is written per unit logarithm of the
argument.  The peak and average densities are related by
\begin{equation}
\rho_{\rm pk}=\rho_{\rm ave}e^{-0.5D^2},
\end{equation}
and the Mach number $M$ and log-normal width $D$ are related approximately by
\citep{pnj97}
\begin{equation}
D^2=\ln(1+0.25{\cal M}^2).\label{eq:D}
\end{equation}
The average density profile is assumed to be a cored power law,
\begin{equation}
\rho_{\rm ave}(r)=\rho_{\rm edge} {{ r_{\rm edge}^\alpha + r_{\rm
core}^\alpha}\over {r^\alpha + r_{\rm
core}^\alpha}}.\label{eq:rho}
\end{equation}
Then equation (\ref{eq1}) becomes \citep{elmegreen11}
\begin{equation}
P^{\prime}_{\rm PDF,3D}(y)= {{3{\cal
C}}\over{\alpha(2\pi)^{0.5}}}\int_{1/{\cal C}}^{1}
\exp\left(-{{\ln^2(yze^{0.5D^2})}\over{2D^2}}\right) {{\left(z{\cal
C}-1\right)^{(3-\alpha)/\alpha}}\over{D\left({\cal
C}-1\right)^{3/\alpha}}}dz,\label{eq:pdftotal}
\end{equation}
where $y=\rho/\rho_{\rm edge}$ is the local density including turbulent fluctuations,
normalized to the value at the cloud edge, $z=\rho_{\rm edge}/\rho_{\rm ave}(r)$ is the
inverse of the average density, and ${\cal C}$ is the degree of central concentration,
\begin{equation}{\cal C}={{\rho_{\rm ave}(r=0)}\over{\rho_{\rm edge}}}.
\label{ceee}
\end{equation}

In the dynamical model of star formation, essentially all of the gas evolves toward
higher density when it is not being pushed back by stellar pressures or sheared out by
galactic rotation. The result is a delayed or {\it resistive} collapse, i.e., one
filled with obstacles, but still a progression toward higher densities at some fraction
of the dynamical rate.  Such a model implies that the 3D star formation rate,
$\rho_{\rm SFR}$, is proportional to the integral of the density-dependent collapse
rate over the entire 3D density PDF:
\begin{equation}
\rho_{\rm SFR}= \epsilon_{\rm ff}\int_0^\infty P_{\rm PDF,3D}(\rho)\left(\rho/t_{\rm ff}\right)d\rho.
\label{sfr1}
\end{equation}
There is no lower limit to the density in this expression because the low-density gas
contributes very little to the integral. In practice, this lower limit is around
$\rho_{\rm edge}$ and the value of that is approximately the average midplane density.
Lower-density gas tends to be warm or hot-phase and unable to join the cooler gas that
is condensing from self-gravity. For numerical integrations in this sub-section, we
take the lower limit of the integral equal to $0.001\rho_{\rm edge}$ and the upper
limit $10^6\rho_{\rm edge}$.

The KS relation uses the SFR surface density, for which we should integrate $\rho_{\rm
SFR}$ over the line of sight through the galaxy. For the moment we write this as
\begin{equation}
\Sigma_{\rm SFR}=2H \rho_{\rm SFR}.
\label{sfr1p5}
\end{equation}

The total gas column density is
\begin{equation}
\Sigma_{\rm gas}=2H \int_0^\infty P_{\rm PDF,3D}(\rho)\rho d\rho
\label{surfdens}
\end{equation}
and the molecular column density is
\begin{equation}
\Sigma_{\rm mol}=2H \int_{\rm \rho_{\rm mol}}^\infty P_{\rm PDF,3D}(\rho)\rho d\rho,
\label{mol1}
\end{equation}
where we use a fixed lower limit to the density in the region of the interstellar
medium where the molecules appear. This is considered to be an effective minimum
density for emission \citep{glover12,shirley15,jimenez17,leroy17a} and is taken to have
a universal value independent of star formation rate and interstellar structure. In
Section \ref{columnthreshold} we also consider a local column density threshold for the
appearance of molecules.

\begin{figure}
\epsscale{.7} \plotone{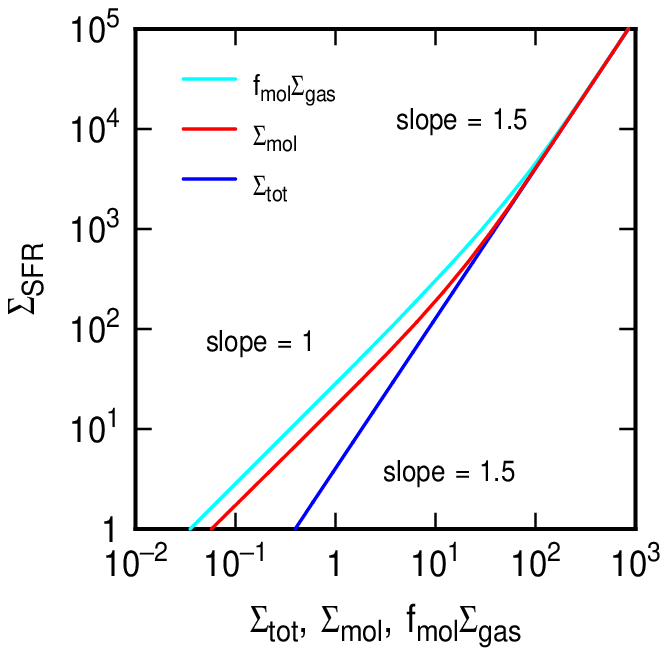}
\caption{Solutions to equations (\ref{eq:pdftotal}), (\ref{sfr1})-
(\ref{fmol}) when the average density profile
inside each cloud has a slope $\alpha=2$. Two parts of the KS law in main
galaxy disks are shown: (1) the 1.5 slope
(blue) for all phases of gas
at high gas surface density when the interstellar medium is mostly
molecular and the same 1.5 slope also for any gas surface density
when plotted versus total gas;
(2) the linear slope for molecules (red) at low-to-intermediate surface
densities when the characteristic density for molecular emission is larger than
the average interstellar density. The red curve uses equation (\ref{mol1}) directly and the
cyan curve calculates the molecular fraction from timing considerations, equation (\ref{fmol}).
The offset between the red line and the cyan line is non-physical, it is the difference between
integrating the PDF over $\rho^{1.5}$ and integrating it over $\rho$ with a separate
multiplication by $\rho^{0.5}$.  The first method is the most physically relevant and
the second illustrates the importance of dynamical evolution in the molecular medium.
}\label{threshold_ks_tot_mol}
\end{figure}

\begin{figure}
\epsscale{.7} \plotone{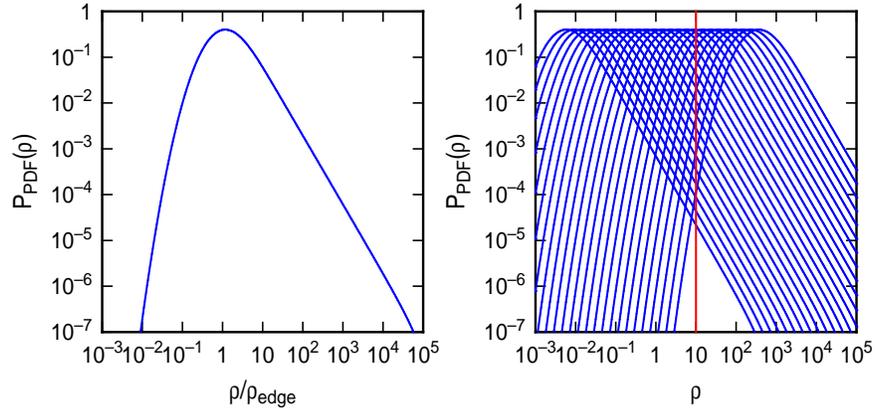}
\caption{(left) The density PDF from equation (\ref{eq:pdftotal}) for a
density contrast ${\cal C}=10^5$, a Mach number 2, and $\alpha=2$, normalized to unit area.
(right) A sequence of PDFs with increasing $\rho_{\rm edge}$ illustrating how the
interstellar PDF moves through a constant threshold for
molecular emission (red line) as $\Sigma_{\rm gas}$ increases. This sequence corresponds
to the points used to plot the curves in Figure \ref{threshold_ks_tot_mol}.
}\label{threshold1}
\end{figure}

Figure \ref{threshold_ks_tot_mol} plots $\Sigma_{\rm SFR}$ versus $\Sigma_{\rm gas}$ as
a blue curve and $\Sigma_{\rm SFR}$ versus $\Sigma_{\rm mol}$ as a red curve assuming
$\alpha=2$. These quantities come from the PDF integrals without conversion to physical
units and are shown to illustrate the slopes. Each curve is a sequence of increasing
$\rho_{\rm edge}$, which tracks the density variation in the interstellar medium. We
assume the scale height $H$ is constant for these curves. The effect of increasing
$\rho_{\rm edge}$ is shown in Figure \ref{threshold1}, which plots the 3D density PDFs
from equation (\ref{eq:pdftotal}) assuming a core-to-edge density contrast ${\cal
C}=10^5$. Each curve uses a different $\rho_{\rm edge}$, increasing from
$7.9\times10^{-3}$ to $3.2\times10^2$ in equation (\ref{eq:pdftotal}) as the curves
move to the right. The vertical line in Figure \ref{threshold1} is the fixed value of
$\rho_{\rm mol}$. As mentioned above, both $\Sigma_{\rm SFR}$ and $\Sigma_{\rm gas}$
come from integrals under the whole curve, but $\Sigma_{\rm mol}$ comes only from the
integral of the part of the curve to the right of the vertical line.

The assumption of a constant scale height in Figure \ref{threshold_ks_tot_mol} makes
the total-gas relation (blue curve) have a power of 1.5 like the 3D relation because
$\Sigma_{\rm gas}$ is proportional to $\rho_{\rm edge}$. The molecular gas relation is
different with a slope of unity. This difference is because of the constant density,
$\rho_{\rm mol}$, used as a lower limit in the PDF integral for the red curve. At high
surface density, the molecular slope also becomes 1.5 because most of the disk has a
density above the molecular excitation density. In terms of the PDFs, this means that
most of the PDF has shifted to the right of the $\rho_{\rm mol}$ in Figure
\ref{threshold1} so the total integral over the PDF from $\rho=0$ to $\rho=\infty$ is
about the same as the integral above $\rho_{\rm mol}$.

This shift from a linear law to a 1.5 law is consistent with observation of the KS law
which show a steepening slope from $\sim1$ in the main parts of spiral galaxy disks to
$\sim1.5$ in the inner regions where the molecular fraction is high
\citep{kennicutt98}. \cite{leroy17b} found a steepening in the relation for CO toward
the inner part of M51. The high $\Sigma_{\rm CO}$ of starburst galaxies in \cite{gao04}
produced a steep molecular relation too.  A steep relation was observed for HCN at high
star formation rates in ULIRGS \citep{gracia08}. \cite{kepley14} observed a high star
formation rate per unit HCN luminosity in the center of the starburst M82. The decrease
in the molecular depletion time with decreasing radius in galaxies found by
\cite{utomo17} may be the same effect.

The cyan curve in Figure \ref{threshold_ks_tot_mol} shows $\Sigma_{\rm SFR}$ versus
$f_{\rm mol}\Sigma_{\rm gas}$, which is the molecular part of the gas determined with
fraction $f_{\rm mol}$ from timing (Paper I). Using the PDF, this fraction becomes the
ratio of average times in the molecular and total-gas phases,
\begin{equation}
f_{\rm mol}={{\int_{\rm \rho_{\rm mol}}^\infty P_{\rm PDF,3D}(G\rho)^{-0.5}d\rho/
\int_{\rm \rho_{\rm mol}}^\infty P_{\rm PDF,3D}d\rho}
\over { \int_0^\infty P_{\rm PDF,3D}(G\rho)^{-0.5}d\rho}/
\int_0^\infty P_{\rm PDF,3D}d\rho};
\label{fmol}
\end{equation}
this is the quantity used in the figure. The cyan curve is parallel to the red curve
throughout Figure \ref{threshold_ks_tot_mol}, meaning that the molecular fraction
obtained from integrating the PDF over densities exceeding $\rho_{\rm mol}$ is the same
as the fraction of the time that the total gas spends in the molecular phase.

\cite{semenov17} obtain the linear molecular law in a numerical simulation because
opacity provides a lower limit to the densities of molecular regions and feedback from
star formation provides an upper limit. With the resulting molecular density in a
narrow range, equations (\ref{mol1}) and (\ref{fmol}) apply and $t_{\rm ff,mol}$ is
nearly constant. Similarly, numerical simulations by \cite{padoan16b} of molecular
cloud formation in a turbulent medium find best agreement with observations when the
molecular fraction increases suddenly at a characteristic density, effectively giving
most of the molecular gas approximately that density.

\begin{figure}
\epsscale{.7} \plotone{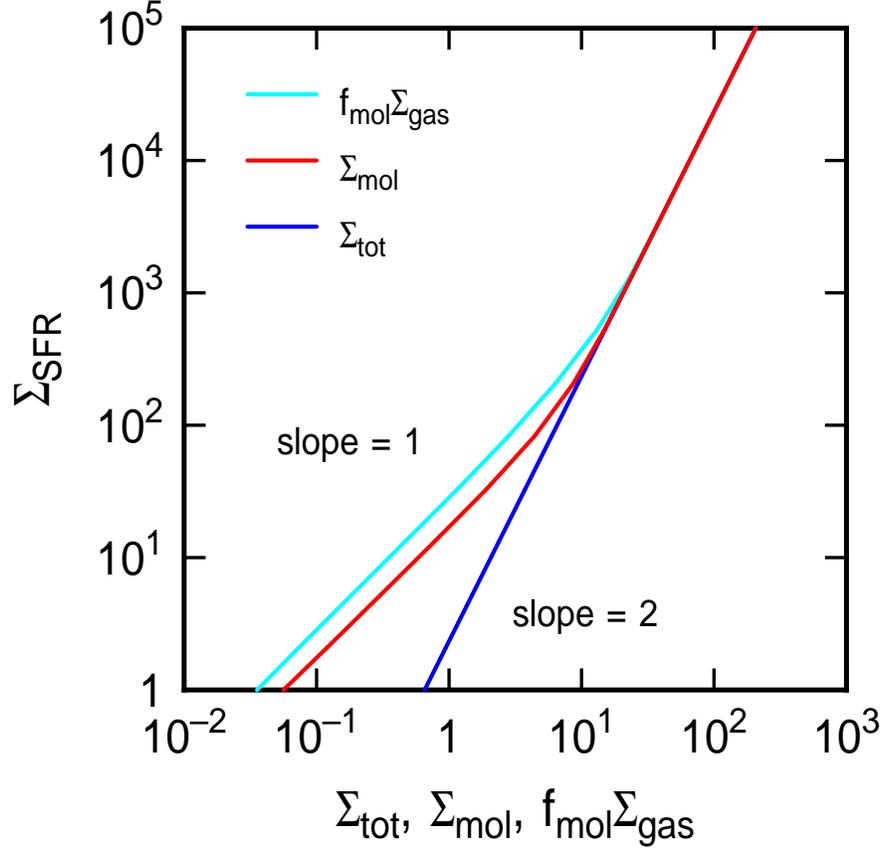}
\caption{Solutions to equations (\ref{eq:pdftotal})-(\ref{fmol}) as in
Figure \ref{threshold_ks_tot_mol} but for a pure-gas disk where the
scale height varies inversely with the gas surface density. Now the
KS relation for total gas (blue) and high-density molecular gas (blue/red) has a slope of
2, although the molecular gas (red) still has a slope of 1 at low
surface density.  $\alpha=2$ is assumed.
}\label{threshold_ks_tot_mol_hvar}
\end{figure}

In gas-dominated regions of a galaxy, such as the outer parts of spirals and most of
dwarf irregulars where $\sigma$ is also nearly constant, the gas scale height is
determined by self-gravity, $H=\sigma^2/\pi G \Sigma_{\rm gas}$, as mentioned above. If
$\rho_{\rm edge}$ is approximately the mid-plane density, then $H=\sigma/\left(2\pi G
\rho_{\rm edge}\right)^{0.5}$, which means that at the edge of the self-gravitating
structures,
\begin{equation}
\rho_{\rm edge}\sim{{\pi G \Sigma_{\rm gas}^2}\over{2\sigma^2}}.\label{rhosigma}
\end{equation}
Figure \ref{threshold_ks_tot_mol_hvar} shows the KS relations in this case, using
equation (\ref{rhosigma}) for $\rho_{\rm edge}$ in equation (\ref{ceee}) for ${\cal
C}$, along with equation (\ref{eq:pdftotal}) for the PDF. Now $\Sigma_{\rm
SFR}\propto\Sigma_{\rm gas}^2$ as in equation (\ref{eq:dwarf}), but $\Sigma_{\rm SFR}$
is still proportional to the first power of $\Sigma_{\rm mol}$ except in the
high-$\Sigma_{\rm gas}$, molecular-dominated regions, where the squared-law appears for
molecules too.

Such a squared-law for molecules has not been observed in galaxies yet because usually
the molecular-dominated interstellar regions are stellar-dominated in mass. Only
individual molecular clouds have shown a squared molecular KS law so far, as discussed
in Section \ref{ks4}. There may be applications of the squared molecular law in
high-redshift galaxies which might have both gas dominating the stars and molecules
dominating the gas \citep[for a review of the high-redshift KS relation,
see][]{tacconi13}.

\subsection{A Simple Model Illustrating the Slopes of the KS Relation in a Dynamical Interstellar Medium}
\label{simple}

Considering the complexity of equation (\ref{eq:pdftotal}), a simpler derivation could
make the results more intuitive. Here we consider just the power-law part of the PDF to
illustrate the various slopes of the KS relations and the effect of widespread density
gradients of the type $\rho\propto r^{-\alpha}$. As mentioned above, this gradient
translates to a power-law column density PDF with a slope of $-1-2/(\alpha-1)$ when
plotted in linear intervals of column density, and $-2/(\alpha-1)$ when plotted in
logarithmic intervals. For 3D density, these PDFs have slopes of $-1-3/\alpha$ and
$-3/\alpha$, respectively. We now use this power law PDF to derive the various KS
slopes shown in Figures \ref{threshold_ks_tot_mol} and \ref{threshold_ks_tot_mol_hvar}.

To be specific, we write the normalized 3D PDF for $\alpha=2$ as
\begin{equation}
P_{\rm 3D,PDF}d\rho=1.5\rho_{\rm edge}^{1.5}\rho^{-2.5}d\rho
\label{simplepdf}
\end{equation}
between $\rho=\rho_{\rm edge}$ and $\rho=\rho_{\rm max}>>\rho_{\rm edge}$; the average
density out to the edge is ${\bar \rho}(r_{\rm edge})=3\rho_{\rm edge}$. We define the
scale height based on the average quantities, indicated by a bar over the symbol:
$H=\sigma\left(2\pi G{\bar\rho}\right)^{-0.5}= \sigma^2\left(\pi G {\bar\Sigma}_{\rm
gas}\right)^{-1}$ where ${\bar\Sigma}_{\rm gas}=2H{\bar\rho}$. Then
\begin{eqnarray}
\label{eq:sfr2}
\Sigma_{\rm SFR}=2H\epsilon_{\rm ff}
\int_{\rho_{\rm edge}}^{\rho_{\rm max}} P_{\rm 3D,PDF}(\rho)\left(\rho/t_{\rm ff}\right)d\rho
=3H \epsilon_{\rm ff}\left(32G/[3\pi]\right)^{0.5}\rho_{\rm edge}^{1.5}
 \int_{\rho_{\rm edge}}^{\rho_{\rm max}}\rho^{-1}d\rho\\
\nonumber
=\left(4G/[9\pi H]\right)^{0.5}\epsilon_{\rm ff}{\bar\Sigma}_{\rm gas}^{1.5}
\ln\left(\rho_{\rm max}/\rho_{\rm edge}\right)\;\;{\rm (constant \;H)}\\
\nonumber
=(2/3)\epsilon_{\rm ff}\left(G{\bar\Sigma}_{\rm gas}^2/\sigma\right)
\ln\left(\rho_{\rm max}/\rho_{\rm edge}\right)\;\;{\rm (constant \;\sigma)}.
\end{eqnarray}
The first result is for a region in a galaxy where the scale height is constant, and
the second result is for a region where the velocity dispersion is constant.  These
expressions have the same forms as equations (\ref{kstotal}) and (\ref{eq:dwarf}),
respectively, with additional weak dependencies on $\rho_{\rm edge}$.

The average surface density comes from an integral as in equation (\ref{surfdens}),
\begin{equation}
{\bar\Sigma}_{\rm gas}=2H \int_{\rho_{\rm edge}}^{\rho_{\rm max}}
P_{\rm PDF,3D}(\rho)\rho d\rho
={{3\sigma\rho_{\rm edge}}\over{\left(2\pi G\right)^{0.5}}}
 \times\int_{\rho_{\rm edge}}^{\rho_{\rm max}}\rho^{-1.5}d\rho
 ={{6\sigma\rho_{\rm edge}^{0.5}}\over{\left(2\pi G\right)^{0.5}}}
\label{surfdens2}
\end{equation}

The average molecular surface density is:
\begin{eqnarray}
\label{surfmol2}
{\bar\Sigma}_{\rm mol}=2H \int_{\rho_{\rm mol}}^{\rho_{\rm max}} P_{\rm PDF,3D}(\rho)\rho d\rho
={{3\sigma\rho_{\rm edge}}\over{\left(2\pi G\right)^{0.5}}}
 \times\int_{\rho_{\rm mol}}^{\rho_{\rm max}}\rho^{-1.5}d\rho\\
\nonumber
 =\left({{1}\over{6H\rho_{\rm mol}}}\right)^{0.5}{\bar\Sigma}_{\rm gas}^{1.5}\;\;({\rm constant\;H})\\
\nonumber
=\left({{\pi G}\over{6\sigma^2\rho_{\rm mol}}}\right)^{0.5}{\bar\Sigma}_{\rm gas}^2\;\;({\rm constant\;\sigma}).
\end{eqnarray}
The $\Sigma_{\rm gas}$ dependencies for molecular column density are the same as the
$\Sigma_{\rm gas}$ dependencies for $\Sigma_{\rm SFR}$, so the two scale linearly with
each other, as discussed in Section \ref{integrals}.

The molecular fraction also follows from this simple model using the fraction of the
time spent in the molecular phase (i.e., at $\rho>\rho_{\rm mol}$), from equation
(\ref{fmol}),
\begin{equation}
f_{\rm mol}={{\int_{\rm \rho_{\rm mol}}^{\rho_{\rm max}} P_{\rm PDF,3D}(G\rho)^{-0.5}d\rho/
\int_{\rm \rho_{\rm mol}}^{\rho_{\rm max}} P_{\rm PDF,3D}d\rho}
\over { \int_{\rho_{\rm edge}}^{\rho_{\rm max}} P_{\rm PDF,3D}(G\rho)^{-0.5}d\rho}/
\int_{\rho_{\rm edge}}^{\rho_{\rm max}} P_{\rm PDF,3D}d\rho}=\left({{\rho_{\rm edge}}\over{\rho_{\rm mol}}} \right)^{0.5}.
\label{fmol2}
\end{equation}
Then $\Sigma_{\rm mol}=f_{\rm mol}\Sigma_{\rm gas}$ from equations (\ref{surfdens2})
and (\ref{surfmol2}), and $\Sigma_{\rm SFR}=\epsilon_{\rm ff}f_{\rm mol}\Sigma_{\rm
gas}/t_{\rm ff,mol}$ to within a factor of $2/\ln(\rho_{\rm max}/\rho_{\rm edge})$.
This slight inaccuracy for the simple model is the difference between integrating over
$\rho^{1.5}$ in equation (\ref{eq:sfr2}) and taking the product of the integral over
$\rho$ from equation (\ref{surfdens2}) and the $\rho$ part of the dynamical rate,
$\rho^{0.5}$.

\begin{figure}
\epsscale{.7} \plotone{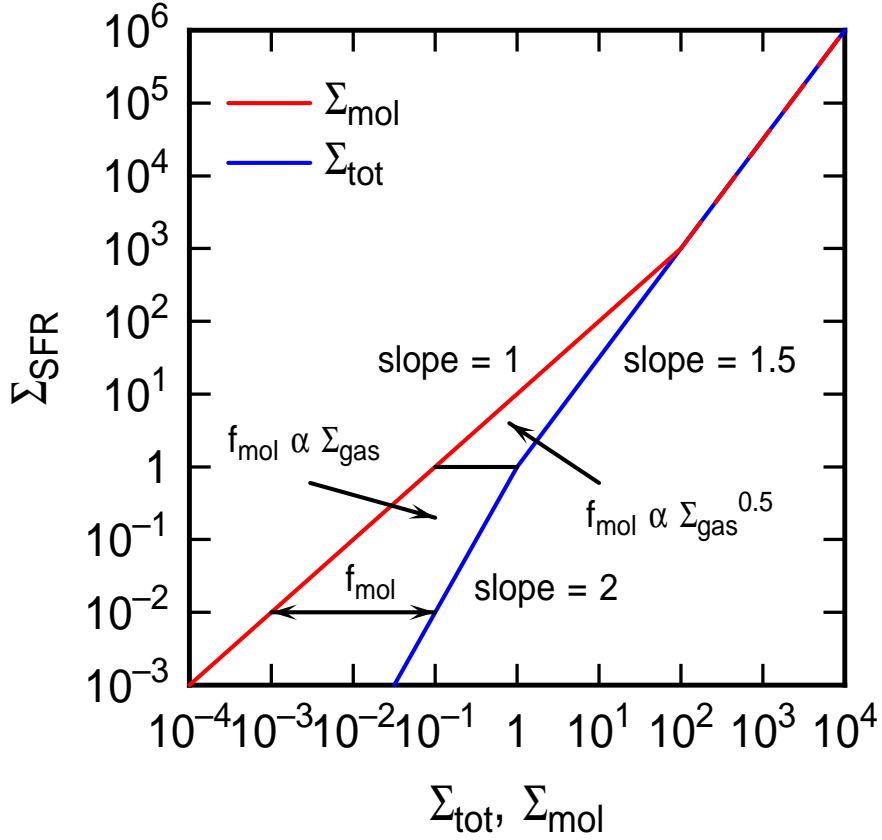}
\caption{Schematic KS relation showing star formation rate surface
density versus total gas (blue) and molecular gas (red)
as calculated from equations (\ref{simplepdf}) -- (\ref{fmol2}), which
assume a power-law PDF.  The slope is 1.5 for all phases at high gas surface
density, 1.5 for total gas at intermediate surface density,
1 for molecular gas at intermediate-to-low surface
density, and 2 for total gas at low surface density. All of these relationships
follow from one three-dimensional star formation law, equation (\ref{sf3d}), but they are viewed
with different radial variations of galaxy thickness and different selection effects.
The molecular fraction is the horizontal distance between the molecular and the
total-gas KS curves in this logarithmic plot.
}\label{threshold_ks_simple}
\end{figure}

The various star formation relations are shown graphically in Figure
\ref{threshold_ks_simple}. The total gas relation is plotted with two blue lines. It
has a slope of 1.5 where the scale height is about constant, which tends to be in the
star-dominated regions at high $\Sigma_{\rm tot}$ and high $\Sigma_{\rm mol}$.  Because
$H\sim$ constant, $\sigma$ is expected to decrease with increasing galacto-centric
radius approximately as an exponential with a scale length that is twice the disk scale
length for the stars. The gas-dominated regions are shown by a blue line at low
$\Sigma_{\rm gas}$ with a slope of 2. In a spiral galaxy, the transition from
star-dominated to gas-dominated occurs in the outer disk, so a single radial profile
should show both KS-1.5 and KS-2a if it goes far enough. In a dwarf irregular galaxies,
only the gas-dominated part might be present and then the total relation is KS-2a with
a slope of $\sim2$, as found by \cite{eh15}. The molecular star formation relation is
shown in Figure \ref{threshold_ks_simple} by a red line. This has a slope of 1 at
low-to-moderate $\Sigma_{\rm mol}$ because of the selection effect to pick regions
defined by a characteristic density, $\rho_{\rm mol}$, which makes $t_{\rm ff}=t_{\rm
ff,mol}$ constant when the molecular fraction is low \citep[e.g.,][]{krumholz07}.

On a log-log plot like Figure \ref{threshold_ks_simple}, the molecular fraction at a
particular star formation rate is the difference between the logs of the molecular and
total surface densities, represented by the horizontal distance between the blue and
red lines. In the star-dominated regions at low-to-moderate $\Sigma_{\rm mol}$, the
molecular fraction scales with the square root of the total gas surface density. In the
gas-dominated regions, it scales with the first power. These scalings are evident
directly from the figure and may also be derived from equation (\ref{fmol2}): if $H=$
constant in the first case, then $\rho_{\rm edge}\propto\Sigma_{\rm gas}$ and $f_{\rm
mol}\propto\Sigma_{\rm gas}^{0.5}$ \citep[as also recognized
by][]{heiderman10,krumholz12}; if $H=\sigma^2/\pi G \Sigma_{\rm gas}$ in the second
case, then $\rho_{\rm edge}\propto\Sigma_{\rm gas}^2$, and $f_{\rm
mol}\propto\Sigma_{\rm gas}$.

The power law expression for the PDF is simple enough to allow us to see a problem with
a SFR based on equation (\ref{ksmol}) if the power in the radial profile of density,
$\alpha$, is not equal to 2, as assumed above.  For a more general case with 3D PDF
power $\gamma=3/\alpha$ on a log-log plot, the normalized PDF is $P_{\rm
PDF,3D}d\rho=\gamma\rho_{\rm edge}^{\gamma} \rho^{-1-\gamma}d\rho$. Then the integrals
in equations (\ref{eq:sfr2})-(\ref{surfmol2}) become
\begin{eqnarray}
\label{eq:sfr3}
\Sigma_{\rm SFR}
=\left({{16G(\gamma-1)^3}\over{3\pi H(\gamma-1.5)^2\gamma}}\right)^{0.5}
\epsilon_{\rm ff}{\bar\Sigma}_{\rm gas}^{1.5}\;\;{\rm (constant \;H)}\\
\nonumber
=\left({{16(\gamma-1)^3}\over{3 (\gamma-1.5)^2\gamma}}\right)^{0.5}
\epsilon_{\rm ff}\left(G{\bar\Sigma}_{\rm gas}^2/\sigma\right)\;\;{\rm (constant \;\sigma)}.
\end{eqnarray}
\begin{equation}
{\bar\Sigma}_{\rm gas}
 =\left({{\gamma}\over{\gamma-1}}\right)^{0.5}\left({{2\sigma}\over{(2\pi G)^{0.5}}}\right)
\rho_{\rm edge}^{0.5}
\label{surfdens3}
\end{equation}
\begin{eqnarray}
\label{surfmol3}
{\bar\Sigma}_{\rm mol}
=\left({{\gamma-1}\over{2\gamma H\rho_{\rm mol}}}\right)^{\gamma-1}{\bar\Sigma_{\rm gas}}^{\gamma}
\;\;({\rm constant\;H})\\
\nonumber
=\left({{(\gamma-1)\pi G}\over{2\gamma \sigma^2\rho_{\rm mol}}}\right)^{\gamma-1}
{\bar\Sigma}_{\rm gas}^{2\gamma-1}\;\;({\rm constant\;\sigma}).
\end{eqnarray}
Now the $\Sigma_{\rm gas}$ dependencies for molecular column density and $\Sigma_{\rm
SFR}$ are not the same if $\alpha\neq2$, so the SFR does not scale linearly with
molecules. The above equations suggest $\Sigma_{\rm SFR}\propto\Sigma_{\rm
mol}^{1.5/\gamma}$, which is sub-linear for shallow cloud profiles, $\alpha<2$, and the
corresponding $\gamma>1.5$. Note that this $\gamma$ is the slope of the 3D density PDF
for log intervals of density, which is not observed directly. The slope of the PDF for
surface density, which is directly observed, is related to this $\gamma$ by
\begin{equation}
\gamma_{\rm 2D}={{2\gamma}\over{3-\gamma}}.
\end{equation}
Thus $\gamma>1.5$ for the slope of the 3D density PDF on a log-log plot corresponds to
$\gamma_{\rm 2D}>2$ for the slope of the observed surface density PDF on a log-log
plot.

This sublinear behavior with $\alpha<2$ and $\gamma_{\rm 2D}>2$ is only for the
molecular KS relationship; the KS slope is still 1.5 for total gas regardless of
$\alpha$ and $\gamma_{\rm 2D}$ (eg. \ref{eq:sfr3}). The sublinear molecular relation
implies that at high gas surface density, an increasing fraction of the observed
molecules are in the form of diffuse clouds, i.e., not strongly self-gravitating, and
therefore not contributing to the SFR at the full dynamical rate. Also, it means that
at low gas surface density, there is star formation in a phase of gas that is not
revealed by that particular molecular emission, in dark molecular gas or atomic gas. We
discuss the sublinear KS relation more in the next section.

\section{The KS Relation with a column density threshold for molecules}
\label{columnthreshold}

Observations suggest that the KS relation for HCN sometimes becomes sublinear with too
little emission for the SFR at high surface densities
\citep{chen15,bigiel16,gowardhan17}, and sometimes becomes sublinear with excess HCN
for the SFR at low surface densities \citep{usero15,kauffmann17a}. Sublinear molecular
emission like this was discussed in the previous section as a possible result of a
shallow density profile inside molecular clouds ($\alpha<2$) in a large-scale survey,
where the projected PDF for the cloud population is steep ($\gamma_{\rm 2D}>2$).

A sublinear slope results when there are molecular emission regions that are not
collapsing into stars, because that moves the observations to the right in the KS
diagram without moving them up. Radiative heating of the HCN by local stars can do this
\citep{shimajiri17}, as that increases the HCN luminosity for a given H$_2$ mass and
star formation rate.  CO observations in local galaxies have also been plotted with a
sublinear slope \citep{shetty14}, with the interpretation that much of the CO is in a
diffuse, non-gravitating phase. Diffuse CO could be responsible for a decrease in the
star formation efficiency with increasing molecular mass for clouds in the Milky Way
\citep{ochsendorf17}. HCN also contains a diffuse component in the central parts of
galaxies, where the pressure is high \citep{helfer97,kauffmann17b}. Molecules that
require a certain column density for self-shielding or extinction, such as H$_2$ or CO
\citep{pineda08}, should have a sub-linear KS relation when this column density is less
than the threshold for strong self-gravity, which depends on pressure (Eq.
\ref{sigmathres}).

A column density threshold for molecule visibility has about the same effect as
decreasing the value of $\rho_{\rm mol}$ for an increase in $\Sigma_{\rm gas}$.
Decreasing $\rho_{\rm mol}$ extends the integral in equation (\ref{mol1}) to include a
larger part of the PDF in molecular form, increasing $\Sigma_{\rm mol}$ without
changing the integral in equations (\ref{sfr1}) and (\ref{eq:sfr2}) that control
$\Sigma_{\rm SFR}$.

In the present model with average density variations like $\rho=\rho_{\rm edge}r_{\rm
edge}^2/r^2$, we can write a characteristic column density $\Sigma_{\rm edge}=\rho_{\rm
edge}r_{\rm edge}$. For a self-gravitating medium at pressure $P$, $\Sigma_{\rm
edge}\sim\Sigma_{\rm Threshold}$ for large-scale $r_{\rm edge}$. Also with this density
gradient, a cloud's mass out to the edge is $M_{\rm edge}=4\pi \rho_{\rm edge} r_{\rm
edge}^3$. Combining these quantities gives
\begin{equation}
r_{\rm edge}=\left({{M_{\rm edge}}\over{4\pi\Sigma_{\rm edge}}}\right)^{0.5}.
\end{equation}
Now consider an effective critical surface density for the appearance of a particular
molecule, $\Sigma_{\rm c}$. Because in general for this density gradient,
$\Sigma(r)=\rho_{\rm edge}r_{\rm edge}^2/r$, the critical radius in a cloud that has
this surface density is
\begin{equation}
r_{\rm c}={{\rho_{\rm edge}r_{\rm edge}^2}\over{\Sigma_{\rm c}}}=\left(
{{\Sigma_{\rm edge}}\over{\Sigma_{\rm c}}}\right)
\left({{M_{\rm edge}}\over{4\pi\Sigma_{\rm edge}}}\right)^{0.5}.
\end{equation}
Putting this radius into the $\rho(r)$ density law gives the corresponding effective
critical density,
\begin{equation}
\rho_{\rm c}={{\Sigma_{\rm edge}r_{\rm edge}}\over{r_{\rm c}^2}}=\Sigma_{\rm c}^2
\left({{4\pi}\over{\Sigma_{\rm edge}M_{\rm edge}}}\right)^{0.5}.
\label{rhoc}
\end{equation}
This equation states that a critical column density has a corresponding critical
density that depends on cloud mass and decreases slowly with increasing pressure
(through Eq. \ref{sigmathres}).

\begin{figure}
\epsscale{.7} \plotone{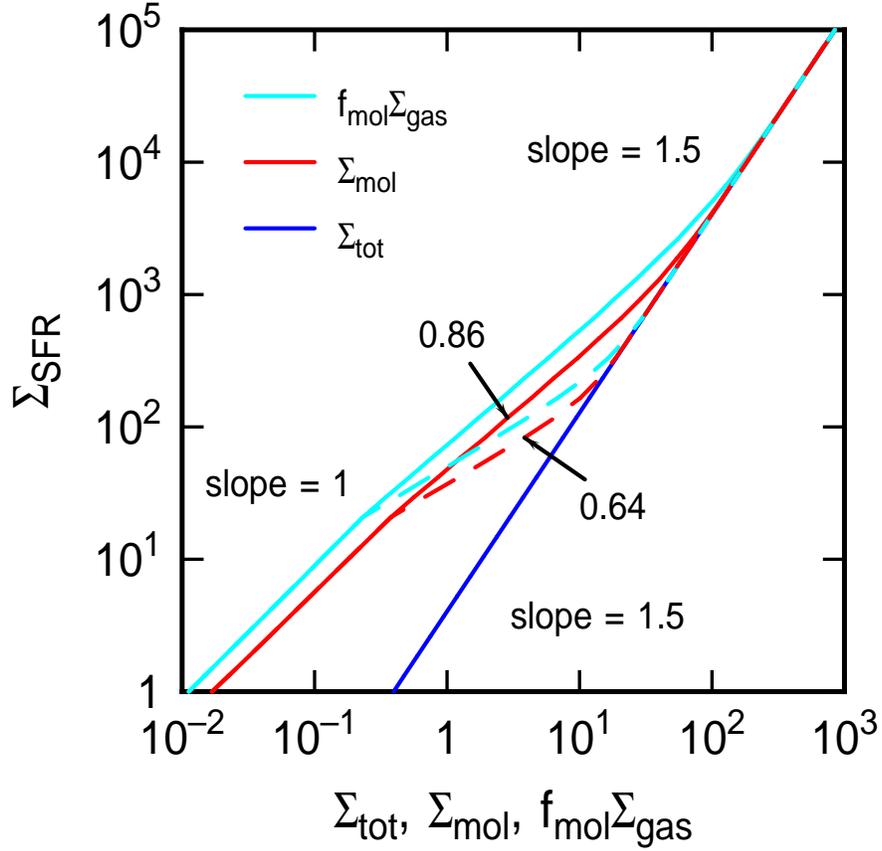}
\caption{Solutions to equations (\ref{sfr1})--(\ref{mol1}) using equation
(\ref{threshold_ks_tot_mol}) as in Figure \ref{threshold_ks_tot_mol},
with the lower limit to equation (\ref{mol1}) taken to be
the critical density $\rho_{\rm c}$ from equation (\ref{rhoc}).  This solution
illustrates the flattening of the KS relation for molecules that have a threshold column
density for formation or emission (which is at $\Sigma\sim0.6$ in this figure).
The split in the red and cyan curves is from two different
models for how the threshold density varies with surface density (see Fig. \ref{threshold_ks_tot_mol}).
}\label{threshold_ks_hcn2}
\end{figure}

Figure \ref{threshold_ks_hcn2} shows the KS relation for a hypothetical molecule that
has a constant density threshold at low $\Sigma_{\rm gas}$, and a constant surface
density threshold at high $\Sigma_{\rm gas}$. The 1.5 slope is still present for the
total gas and for the molecular emission at high surface density, the linear slope is
present for the molecule at low surface density, but now a sub-linear slope is present
at moderate-to-high surface density where there is a surface density threshold for the
molecule. Two sample cases are indicated by the split in the red and cyan lines. These
curves are solutions to equations (\ref{sfr1})--(\ref{mol1}) using equation
(\ref{eq:pdftotal}) as in figure \ref{threshold_ks_tot_mol}, but now the lower limit to
equation (\ref{mol1}) is $\rho_{\rm c}$ from equation (\ref{rhoc}). $\rho_{\rm c}$ is
taken to be 100 for the same range of $\rho_{\rm edge}$ as in Figure \ref{threshold1},
but for $\rho_{\rm edge}>1$, $\rho_{\rm c}$ in equation (\ref{mol1}) is replaced by
$\rho_{\rm c}=100\rho_{\rm edge}^{-0.5}$ in one case (slope 0.86 line) and
$100\rho_{\rm edge}^{-2}$  in the other case (slope 0.64 line).

These hypothetical examples illustrate how molecular tracers can have a sublinear slope
in the KS relation if the molecular gas is in a diffuse, non-self-gravitating state.
Such a state tends to occur when the cloud column density is less than the threshold
value given by equation (\ref{sigmathres}). If the column density of a molecule exceeds
a first threshold for the existence or appearance of the molecule, but not the second
threshold given by self-gravity (Eq. \ref{sigmathres}), then it should present a
sub-linear slope on the KS relation. Another way to say this is that molecular clouds
of a certain column density (for detection) that are self-gravitating at low pressure
will not be self-gravitating at high pressure.

At high column density, i.e., one that is above both the detection and the self-gravity
thresholds, the KS relation for total gas mass should be recovered, i.e., with a slope
of 1.5 if the galaxy thickness is about constant. This is shown in Figure
\ref{threshold_ks_hcn2} as well. The sub-linear to super-linear transition is
reminiscent of observations by \cite{leroy17b}, who find this for CO in M51.

\cite{mok16} found that the molecular fraction, as determined from CO(3-2) and HI
emission, for gas in Virgo spiral galaxies is higher than in group galaxies. They also
found that the molecular emission is high in Virgo compared to the star formation rate.
These observations suggest there is an excess of non-star-forming, or diffuse, CO gas
in Virgo spirals compared to group spirals. According to equations (\ref{sigmathres})
and (\ref{rhoc}), a high interstellar pressure would do this by lowering the density at
which CO is observed, thereby making proportionally more CO, and by increasing the
column density at which clouds become self-gravitating, thereby limiting the fraction
of the interstellar medium that forms stars. High pressure in Virgo cluster spirals is
expected because of the high intergalactic pressure from hot gas and the ram pressure
from galaxy motions \citep{mok17}.

\section{The appearance of threshold densities and column densities for star formation}
\label{threshold}

\subsection{Observations and Models with Thresholds}

In typical regions that have been observed, stars tend to form where the column density
exceeds a certain value corresponding to $\sim8$ mag of visual extinction in the solar
neighborhood
\citep{onishi98,goldsmith08,lada10,heiderman10,froebrich10,kainulainen11,schneider13,evans14},
and they also form at a rate approximately proportional to the mass of dense gas as
traced by HCN \citep{gao04,wu05,wu10,zhang14,shimajiri17}, CS \citep{wang11},
far-infrared emission \citep{vuti14,vuti16}, and extinction
\citep{lada10,heiderman10,lada12,evans14}.

These observations have led to the idea that star formation occurs at densities or
surface densities higher than some threshold value, and that it occurs at a nearly
constant rate above that value, regardless of density. This section discusses the first
point, the appearance of a threshold density, using PDFs for collapse timing. The next
section applies the result to dense gas surveys as an illustration of the second point.

Threshold densities have been part of star formation theory for a long time. They are
assumed in numerical simulations to prevent excessively short timesteps that slow the
evolution down. For example, the EAGLE simulation \citep{schaye15} has a threshold
density that varies with metallicity from galaxy to galaxy as $0.1(Z/0.002)^{-0.64}$
cm$^{-3}$, and Illustris \citep{vogelsberger14} has a constant threshold, 0.13
cm$^{-3}$. \cite{hu16} simulate dIrrs with a constant threshold of $100$ cm$^{-3}$.
\cite{hopkins17} note that the threshold value does not matter in a simulation as long
as it is high enough to avoid the essential physics of interstellar collapse; this is
the same point as in the present paper, where there is no physical density threshold
separating star-forming gas from non-star-forming gas \citep[see also][]{saitoh08}.

Analytical derivations of the star formation rate
\citep[e.g.,][]{elmegreen02,krumholz05,hennebelle11,padoan11} also assume threshold
densities to get the efficiency of star formation correctly when integrating over the
PDF above this density. For example, \cite{padoan17} define a critical density as the
external density of a critical Bonnor–-Ebert sphere that fits in the postshock layer of
a supersonically turbulent gas.  \cite{hennebelle11} choose a critical density at which
fluctuations smaller than a fixed fraction of the cloud size can fragment. Various
models like this are reviewed in \cite{fed12} and \cite{padoan14}.

The dynamical model discussed here does not need a threshold density for star formation
because it reproduces the large-scale star formation properties of galaxies by assuming
the entire interstellar medium is evolving at some fixed fraction of the dynamical
rate, with no transition below and above any particular density. The model only has
characteristic densities or surface densities for the detection of certain molecules,
but not for the star formation process itself.  The same model was applied on the
molecular cloud scale by \cite{parmentier13} to study the formation of bound clusters.
\cite{parmentier16} also considered there is no physical density threshold for star
formation, but interpret the appearance of one as the result of combining local sources
with a steep KS relation inside individual clouds and distant sources with a linear
relation from poor angular resolution. \cite{burkert13} and \cite{lada13} explain the
appearance of a threshold as the result of a decreasing cloud area at higher surface
density, with no actual threshold-like change for the KS relation inside molecular
clouds.

Threshold-free models do not deny that certain densities play an important role in
regulating the various stages of star formation. At high density, magnetic fields
should decouple fast from the gas \citep{goodman98,elmegreen07b} and turbulent motions
become subsonic causing turbulent fragmentation to stop
\citep{padoan95,vazquez03,krumholz05}. The present model implies two other things
instead, that the rate limiting step for star formation on a large scale is the free
fall time at the lowest density, and that stars of a certain young age tend to appear
where $t_{\rm ff}$ is about this age. Also, the lack of a bump or leveling off of the
density PDF in star forming regions prior to the power law part from collapse implies
that there is no bottleneck at some physical threshold between no star formation and
star formation.

\subsection{Star Formation in Dense Gas as a Probabilistic Effect rather than a Threshold}

Here we show that the appearance of a threshold density for star formation could be a
manifestation of the dynamical process itself in the sense that very young objects tend
to appear at a density where the dynamical time is comparable to their age. Star
formation tracers with very young ages tend to show up at densities where the dynamical
time is short.

This proposed correlation between stellar age and dynamical time follows from two
concepts. First, stars drift a distance proportional to their age and relative speed,
so young stars are still near their birth sites, and, second, interstellar gas tends to
be hierarchically clumped, so if a star forms in a certain dense region, then there is
likely to be another dense region nearby \citep[e.g.,][]{gouliermis17,grasha17}.
Putting these together means that young stars tend to be found near high gas densities,
and slightly older stars, which have drifted further from their birth sites, tend to be
near gas that has a more average, i.e., lower, density. Stars sufficiently old will
have drifted past many molecular clouds in their lifetimes and they will have lost any
correlation with the density of their birth.

The implications of this proposed age-density correlation may be illustrated by the
time-dependent collapse of a spherical cloud core, whose radius evolution is given by
\citep{spitzer78}
\begin{equation}
{{d^2r}\over{dt^2}}=-{{GM(r_{\rm init})}\over{r^2}}.
\label{collapsesol}
\end{equation}
The solution is
\begin{equation}
r=r_{\rm init}\cos^2\beta
\end{equation}
where
\begin{equation}
\beta+0.5\sin 2\beta={{\pi}\over 2}{{t}\over{t_{\rm ff}}}
\end{equation}
is solved iteratively. Here, $t$ is the time when $r=r_{\rm init}$ and $t_{\rm ff}$ is
the free fall time at the density when the collapse begins. The solution gives the
radius as a function of time and therefore the density as a function of time.

\begin{figure}
\epsscale{.7} \plotone{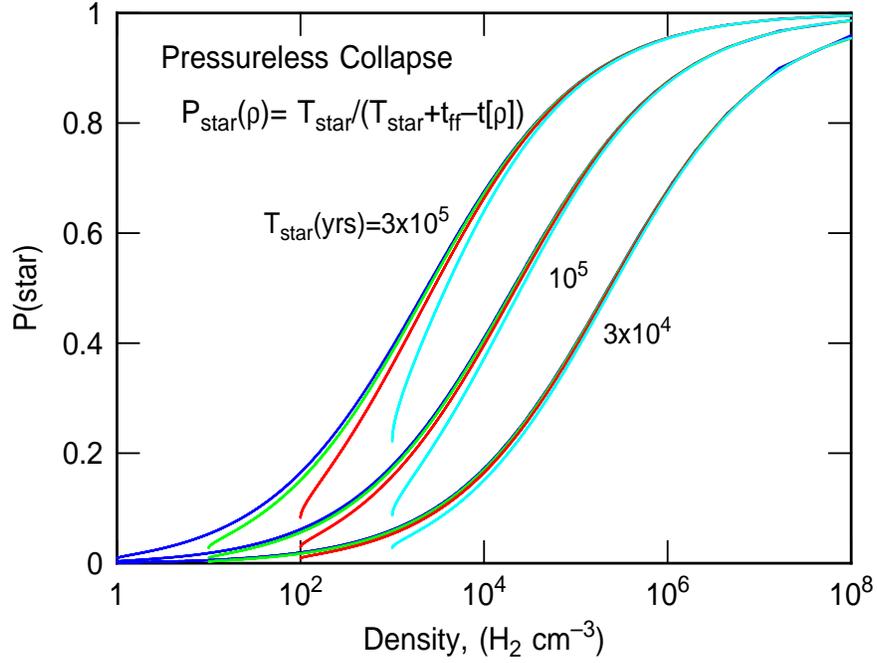}
\caption{Solutions to equations (\ref{collapsesol})--(\ref{eq:prob}) for the probability of
observing stars with a young age, $T_{\rm star}$, near a region of a cloud with the H$_2$
density indicated on the abscissa. The probability increases rapidly at the density
where the free fall time is comparable to the age of the star. Different curves for the same
$T_{\rm star}$ are for
different starting densities in the collapsing cloud.
}\label{lund_collapse}
\end{figure}

If stars form hierarchically (i.e., together) on a dynamical timescale, then the
youngest stars will still be near gas where other stars are forming on about the same
timescale. If young stars of a particular type have an average age $T_{\rm star}$, then
these stars are likely to be in a region where the time before other star formation,
i.e., at the singularity of the collapse, $t-t_{\rm ff}$, is comparable to $T_{\rm
star}$. The density at this time $t$ is known from the above solution, so we can plot
the probability of seeing this star, which is approximately
\begin{equation}
P_{\rm star}(\rho)={{T_{\rm star}}\over{T_{\rm star}+t_{\rm ff}-t(\rho)}},
\label{eq:prob}
\end{equation}
as a function of the instantaneous density, $\rho(t)$. This plot is shown in figure
\ref{lund_collapse} for 3 stellar ages and 4 initial densities of the collapse, 1, 10,
$10^2$, and $10^3$ in cm$^{-3}$ of H$_2$; Helium and heavy elements are included in the
mass density for $t_{\rm ff}$. The different initial densities show up as different
starting points for the curves in the lower left, but at the low values assumed, they
have little effect on $\rho(t)$ after the start.

The curves show an increasing probability of seeing the stars with increasing density.
The time at the half-probability point, in units of the starting $t_{\rm ff}$, ranges
from 0.999 to 0.991 for a starting density of 1 H$_2$ cm$^{-3}$ as $T_{\rm star}$
increases from $3\times10^4$ yrs to $3\times10^5$ yrs, it ranges from 0.997 to 0.972 at
a starting density of 10, 0.991 to 0.910 at a starting density of 100, and 0.972 to
0.716 at a starting density of 1000 H$_2$ cm$^{-3}$. The stars appear at the very last
few percent of the collapse time after the density has become high, regardless of when
the collapse started. Such timing would give the appearance of a threshold density when
in fact the collapse is continuous with no physical threshold separating stability from
collapse. Similar curves result from other collapse solutions \citep[e.g.,][]{huff06}
as long as there is a singularity in density at a certain time.

Although this is an idealized model of singular cloud core collapse, it illustrates the
basic point that in a probabilistic sense, the youngest objects should appear near gas
at the highest densities.  They should also appear near each other because of the
hierarchical structure of the gas, and those that have already formed should be near
others that are just about to form. This correlation between age and density should
persist even to larger scales, as long as the interstellar medium is in a state of
resisted collapse where all phases last for some relatively constant number of
dynamical times within the full cycle of cloud formation and dispersal.

\section{On the appearance of a constant star formation rate per unit dense gas mass}
\label{constant}

The increased likelihood for young stars to appear in dense gas implies that star
formation rates correlate best with the mass of dense nearby gas. Such correlations are
commonly found, and they seem to contradict the dynamical model which also involves
density \citep[e.g.,][]{evans14}. Here we show that the star formation rate should be
weakly dependent on density, as observed, even in the dynamical model if dense gas is
defined either by emission from molecules with a fixed characteristic density for
emission, or by a high column density observed in dust emission or extinction. The
first point with dense gas defined by dense-tracing molecules like HCN is essentially
the same as that discussed in Sections \ref{sfrinhcn} and \ref{pdfsection}, where the
density used for the free fall time is a factor of order unity times the characteristic
density of the molecule's emission.

The second point follows from equation (\ref{rhoc}) when a region is defined by a
threshold column density, $\Sigma_{\rm c}$. In Section \ref{columnthreshold} we used
this equation to suggest that the KS relation flattens for molecular column densities
that exceed a threshold for excitation but not a threshold for self-gravity, given the
ambient turbulent pressure. However, this flattening of the KS relation is also the
main point of the observation that the star formation rate per unit mass is constant,
regardless of density. Equation (\ref{rhoc}) shows that the threshold density for
observation is insensitive to the actual cloud column density or mass when there is a
constant value, $\Sigma_{\rm c}$, used to define the ``dense'' region.  Dense gas
surveys that define their selection to have extinctions exceeding 8 mag or some such
value are in this category. The free fall rate in these regions scales with $\rho_{\rm
c}^{0.5}$, which scales with $\Sigma_{\rm edge}^{-0.25}$ from equation (\ref{rhoc}),
which, at the threshold of self-gravity, scales with interstellar pressure, $P$, as
$P^{-0.125}$ according to equation (\ref{sigmathres}). Thus there is a very weak
dependence on density or ambient conditions once a self-gravitating region is chosen to
exceed a certain fixed column density. The appearance of a fixed star formation rate
per unit dense gas mass is thus a selection effect resulting from the definition of
dense gas.

Setting the threshold column density at 8 magnitudes of visual extinction, which
corresponds to $\Sigma_{\rm gas,threshold}=160\;M_\odot$ pc$^{-2}$, we can rewrite
equation (\ref{sigmathres}) as
\begin{equation}
P_{\rm threshold}=8.8\times10^5k_{\rm B}\left({{\Sigma_{\rm gas,threshold}}\over
{160\;M_\odot\;{\rm pc}^{-2}}}\right)^2.
\end{equation}
The corresponding average molecular density for a velocity dispersion of $\sigma_{\rm
threshold}=10$ km s$^{-1}$ is $n_{\rm H_2}=P/(\mu\sigma^2)=30$ cm$^{-3}$ for mean
molecular weight $\mu$. Evidently, when the average interstellar density exceeds the
equivalent of 30 H$_2$ cm$^{-3}$ at a 10 km s$^{-1}$ velocity dispersion, ``dense''
regions with more than 8 magnitudes of extinction start to become diffuse and should
lose their ability to form stars.  This is another way of saying that the virial
parameter, $5R\sigma^2/(GM)$ becomes large at the corresponding cloud radius $R$ and
mass $M$. Evidently, at very high pressures, ``dense'' gas in star forming regions
should not be defined by 8 magnitudes of extinction. Defining dense gas in terms HCN
emission or other dense molecular tracers may still be practical, but then the linear
relation follows (KS-1a) until the average interstellar density is higher than the HCN
density. \cite{gowardhan17} recognized this problem with the standard definition of
dense gas and suggested that density contrast rather than absolute density is
important.

To be more quantitative, we consider the star formation rate per unit dense gas mass,
which is conveniently defined in terms of the dense gas depletion time, $M_{\rm
dense}\left(dM_{\rm stars}/dt\right)^{-1}$ for dense gas mass $M_{\rm dense}$ and star
formation rate $dM_{\rm stars}/dt$. For surveys with HCN and high density molecular
tracers, this time has been evaluated to be $\sim56$ Myr for galaxies in \cite{gao04},
$\sim69$ Myr for normal spiral galaxies and $\sim37$ Myr for ULIRGS in \cite{liu15},
$\sim83$ Myr for local clouds and galaxies combined in \cite{wu05} and \cite{retes17},
and $\sim60$ Myr for local clouds in \cite{heiderman10}. For surveys at high extinction
or in cloud regions with high FIR emission intensities, the dense gas depletion time
has been evaluated as $\sim22$ Myr \citep{lada12,shimajiri17} and $\sim25$ Myr
\citep{evans14} for local molecular clouds. For a combination of local clouds and
galaxies using both $^{13}CO$ and FIR emission, \cite{vuti16} got $\sim66$ Myr.

These depletion time vary between $\sim20$ Myr and $\sim80$ Myr, depending on the
sources and the observational techniques. For a representative value of $\sim50$ Myr,
which means
\begin{equation}
{{dM_{\rm stars}}\over{dt}}={{M_{\rm dense}}\over{50\;{\rm Myr}}}=\epsilon_{\rm ff}
M_{\rm dense} /t_{\rm ff,dense}
\end{equation}
we require $t_{\rm ff,dense}/\epsilon_{\rm ff,dense}\sim50$ Myr. With a characteristic
density for these regions of $n_{\rm dense}\sim3\times10^4$ cm$^{-3}$, $t_{\rm
dense}=0.19$ Myr and $\epsilon_{\rm ff}=0.004$.  This is about the same result as in
equation (\ref{eq:total2}).

Section \ref{sfrinhcn} discussed how some observed correlations between dense gas
fractions and star formation rates can be explained in an approximate fashion by the
present model. A more fundamental observation is the dense gas fraction itself. For
pervasive density gradients like those assumed here, $\rho\propto1/r^2$, the mass of a
gas concentration increases linearly with size and the column density decreases
linearly with the inverse of size. This implies that the mass ratio of high density to
low density gas in a cloud scales inversely with the column density thresholds used to
define these gases. If extinction $A_{\rm V}=2$ defines the low density part and
$A_{\rm V}=8$ defines the high density part \citep[e.g.,][]{evans14}, then $f_{\rm
dense}=25$\%.  If dense gas is defined by line emission that is sensitive to density
rather than column density, then because mass out to some cloud radius scales inversely
with the square root of density, $f_{\rm dense}$ should be the square root of the ratio
of densities of the low- and high-density tracers, which might be
$\sqrt{0.01}\sim10$\%. These fractions are higher than the observed dense gas fraction
on a large scale, which is more like 4\% \citep{gao04,gowardhan17}, but the large scale
contains other gas not related to star formation. Still, the pervasive appearance of
self-gravitating structures shown by the power-law density PDFs on galactic scales
suggests that the dense gas fraction should vary only slowly with environment, such as
with radius in a galaxy or from galaxy to galaxy where the average density is less than
the characteristic density of the high density tracer (e.g., normal star-forming
galaxies, rather than ULIRGs). In fact, \cite{ragan16} find $f_{\rm dense}\sim$
constant with radius in the Milky Way even though the star formation fraction in this
dense gas varies slightly.

The dense gas fraction defined by line emission, e.g., $L_{\rm HCN}/L_{\rm CO}$, can
increase substantially in ULIRGS and other regions where the average density exceeds
the threshold for CO emission. The reason for this was given in Section
\ref{section:sfe}, i.e., the HCN luminosity increases linearly with the star formation
rate when $f_{\rm mol,HCN}<1$, and the CO luminosity increases with the 2/3 power of
the star formation rate when $f_{\rm mol,CO}\sim1$. Then the ratio of HCN to CO
luminosity increases with the 1/3 power of the star formation rate, making it seem like
regions of high star formation are different with larger dense gas fractions. In fact
it is only a comparison of $f_{\rm mol}<1$ transitions with $f_{\rm mol}\sim1$
transitions that does this. At low star formation rates, $f_{\rm mol}<1$ for both high
and low density molecules and both scale linearly with the star formation rate, giving
a constant dense gas fraction. The dense gas fraction is defined by an observable and
the observable has selection effects that influence the KS relation as much as any
detail of the underlying star formation law.

\begin{figure}
\epsscale{.7} \plotone{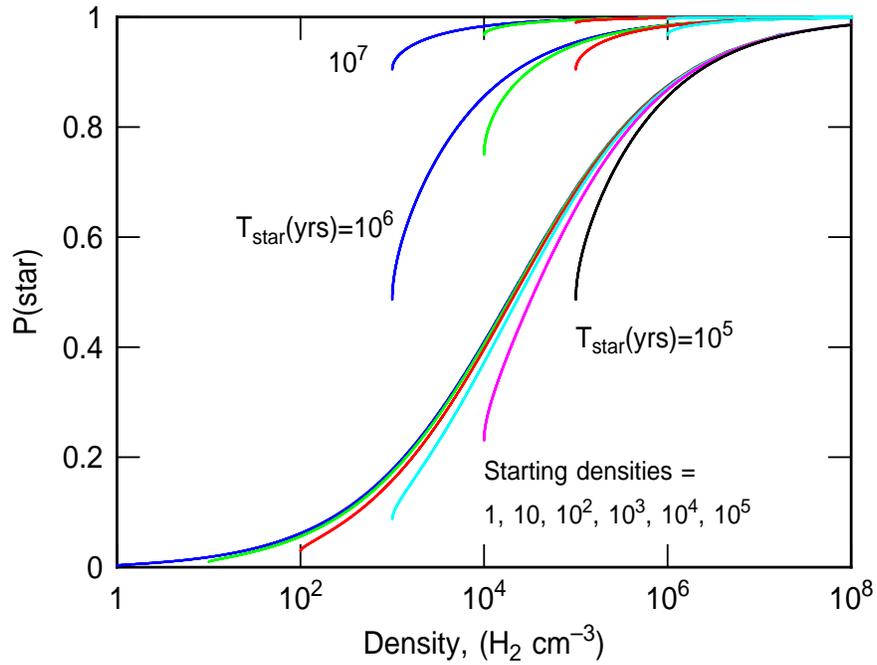}
\caption{Same as Figure \ref{lund_collapse} but for a wider range of starting densities in the
lower curves and for larger stellar ages in the upper curves. This figure shows that
stars slightly older than the free fall time in a gas selected for observation will
usually appear to be associated with that gas, giving the impression for the youngest
observable stars that they tend to form where the {\it average} gas density is high.
}\label{lund_collapse2}
\end{figure}

Now consider the collapse model from Section \ref{threshold}. In a region where $t_{\rm
ff}<T_{\rm star}$, such as the dense part of a molecular cloud or a starburst galaxy
with a high average interstellar density, the curves in Figure \ref{lund_collapse} all
lie at high probability. Figure \ref{lund_collapse2} shows examples. Now the starting
densities for the $T_{\rm star}=10^5$ yrs curves include higher values, $10^4$
cm$^{-3}$, $10^5$ cm$^{-3}$ and $10^6$ cm$^{-3}$, representing surveys of dense-gas
regions inside molecular clouds. Also, for observations of distant galaxies, only older
ages for young stellar regions can be discerned because individual stars and protostars
often cannot be seen by themselves. For example, $T_{\rm star}$ might be as high as 1
Myr for the brightest young stars, or 10 Myr for stars that excite HII regions. These
cases are also in Figure \ref{lund_collapse2}. For high interstellar densities or
relatively old stars at modest densities, the probability of seeing a star associated
with the gas tends to be high.  This would lead to the impression that stars form
primarily where the average density is high, with lower average densities for older
stars. The spatial and temporal resolution of events on short timescales is lost when
equally young stars are not distinguished.

\section{Discussion}
\label{discussion}

A physical model for star formation that is consistent with the above explanation for
the KS relations is discussed here. The main assumption is that gaseous gravity is the
primary driver of star formation on a scale comparable to the scale height. As is well
known, interstellar gas in a disk galaxy is stabilized against gravitational collapse
by rotation on large scales and pressure on small scales
\citep{safronov60,toomre64,goldreich65}. The mass surface density, $\Sigma_{\rm gas}$,
and epicyclic frequency, $\kappa$, determine the rotationally stabilized length in the
radial direction, which is the inverse Toomre wavenumber $k_{\rm T}^{-1}=2\pi
G\Sigma_{\rm gas}/\kappa^2$, where self-gravity balances the Coriolis force. Because a
region of this size contains the mass $M_{\rm T}=\pi\Sigma_{\rm gas} k_{\rm T}^{-2}$,
it has a specific potential energy $\sigma^2/2$ equal to $GM_{\rm T}k_{\rm T}$, which
corresponds to a potential velocity dispersion $\sigma=2\pi G\Sigma_{\rm gas}/\kappa$.
For conditions in the solar neighborhood, $\Sigma_{\rm gas}\sim10\;M_\odot$ pc$^{-2}$
and $\kappa=0.037$ Myr$^{-1}$, this dispersion is $\sigma=7.5$ km s$^{-1}$, much larger
than the sound speed in the cool, diffuse phase of interstellar gas, which is only
$\sigma_{\rm s}=0.8$ km s$^{-1}$ at 100 K. The result should be
gravitationally-generated motion that is supersonic, compressive, and dissipative, with
a tendency to collapse and form stars.

Magnetic forces, galactic shear, and intermittent local expansions from supernovae and
other stellar feedback act to resist this collapse
\citep{kim15,padoan16a,pan16,ibanez16,maclow17,semenov17}. Kinematic pressures like
these stabilize a disk on scales smaller than the Jeans length, $k_{\rm
J}^{-1}=\sigma^2/\pi G \Sigma$. For an adiabatic gas, conventional theory states that
if the Jeans length is larger than the Toomre length, corresponding to the condition
$Q^2=2k_{\rm T}/k_{\rm J}>1$, then there should be stability in the radial direction of
a galaxy. Also, if $Q_{\rm eff}>1$ for $Q_{\rm eff}^{-2}\sim Q^{-2}+2(k_{\rm
azim}\sigma Q/\kappa)^2$, then both radial and azimuthal motions at wavenumber $k_{\rm
azim}$ are stabilized \citep{lau78,bertin89}. However, kinematic resistance at
supersonic speed damps quickly \citep{stone98,maclow98}, so the Jeans length evaluated
with supersonic $\sigma$ should not be a short-term barrier to collapse on small scales
\citep{elmegreen11}. In that case, $Q$ is not a measure of absolute stability. For
example, in low surface-brightness regions where $Q$ appears to be high, star formation
still seems normal \citep{eh15}.\footnote{Observations of disk quenching
\citep{martig09,genzel14,french15} seem to contradict these statements, but perhaps
quenching is from other effects, such as a high supernova rate compared to the average
collapse rate, or high shear \citep{davis14}.} Moreover, a combined $Q$ from gas and
stars appears to be dominated by stellar mass and motions \citep{romeo17}, which means
that stellar spirals, when they are present, do most of the work to regulate $Q\sim2$,
not gas processes such as star-formation feedback. Stars form because self-gravity,
pressure gradients, magnetic torques, and energy dissipation decouple the gas from the
stars on scales less than $k_{\rm T}^{-1}$. This makes gravitationally-driven
contraction of the gas inevitable for a wide range of scales
\citep[e.g.,][]{kravtsov03,li05,ballesteros11,ibanez17}.

The role of the background stellar disk in the star formation process could be minor as
long as the gas can dissipate thermal and turbulent energy. Gas that forms new stars
has to move through the background stars to come to a high density. The two-fluid
instability does not include this process because it assumes that the two fluids have
perturbations with the same wavelength, which is usually the large scale of spiral
arms. Gas moves through stars because of pressure forces from turbulence and stellar
feedback, and it decouples from stars during the gravitational swing amplifier because
of magnetic forces \citep{elmegreen87,kim01}. The point of view here is that background
stars mostly affect the thickness of the gas layer through gravity, heating, and the
generation of turbulence \citep[e.g.,][]{ostriker10} and therefore it affects the
conversion between the observed surface density and the spatial density that controls
the collapse rate. This background effect is important in the main disks of spiral
galaxies and possibly elsewhere, but in the outer parts of these disks or in dwarf
Irregular galaxies, the gas surface density dominates the stellar surface density and
although starlight continues to heat the gas, a disk flare has a larger effect on the
KS relation than the thermal properties \citep{eh15}. As long as there is a cool phase
with a thermal Jean length less than the Toomre length, and as long as the timescales
for thermal, turbulent, and feedback processes are comparable to or shorter than the
timescale for gravitational processes, gravity, as a persistent and monotonic force
toward condensation, should control the rate of star formation.  Then the preceding
derivations of the KS relation should contain most of the relevant physics.

\section{Conclusions}
\label{conclusions}

Galaxies form stars because feedback and other energy sources are unable to resist
self-gravity when gas motions are supersonic and highly dissipative. The resulting
collapse is not rapid, however, because that would mix the young stellar populations
and erase widespread correlations between positions and ages, which suggest turbulent
compression is involved.  The collapse is more likely resistive, with a rate
proportional to the gravity rate but significantly slower. In this case, the star
formation rate per unit volume is related to the gas density in a fairly simple way
(Eq. \ref{sf3d}), and the density structure in the gas may also be written simply using
a power-law PDF.

With these two assumptions, the various relationships between the star formation rate
and the amount of gas can be explained by a combination of resistive collapse and a
selection effect for gas observations.  These relationships were divided into three
types: (KS-1.5) $\Sigma_{\rm SFR}\propto\Sigma_{\rm total\;gas}^{1.5}$ for main galaxy
disks with constant thickness; (KS-1a) $\Sigma_{\rm SFR}\propto\Sigma_{\rm mol}$ for
the molecular part of the gas when the average density is less than the characteristic
density for observation of the molecule (and KS-1.5 again when the average density
exceeds this characteristic density); (KS-1b) $\Sigma_{\rm SFR}\propto\Sigma_{\rm
gas,dense}$ and (KS-1c) $dM_{\rm star}/dt\propto M_{\rm dense}$ for dense gas when the
selection of this gas involves an observational threshold for either density or column
density; (KS-2a) $\Sigma_{\rm SFR}\propto\Sigma_{\rm total\;gas}^2$ for galaxies or
parts of galaxies where gaseous self-gravity determines the disk thickness and the
velocity dispersion is about constant (e.g., outer parts of spirals and dIrrs); and
(KS-2b) $\Sigma_{\rm SFR}\propto\Sigma_{\rm gas}^2$ inside molecular clouds where the
size is also determined by gaseous self-gravity.

Numerous observations were collected to illustrate these four regimes, and the assumed
model for star formation was shown to agree with the observations fairly well if the
efficiency per unit free fall time is always around one percent. The model ignores the
details of collapse and feedback, but the general agreement with these observations
lends support to the basic premise that star formation is a pervasive dynamical process
on all relevant scales.

The star formation relationships were also reproduced in general form using a
probability distribution function for interstellar gas that results from turbulence
convolved with self-gravitational density gradients. Using these PDFs, the molecular
component can be determined either from integration above the characteristic density
for observation of the molecule or from averaging over the time spent in the molecular
phase. These two methods are equivalent for the dynamical model (Paper I).

A column density threshold for the detection of certain molecules flattens the KS
relations by increasing the proportion of diffuse (non-gravitating) gas as the average
interstellar surface density increases. Such flattening has been observed for the HCN
KS relationship at both low and high pressure in different situations. Cloud selection
at a minimum column density (e.g., using a lower limit to the extinction or
far-infrared intensity) leads to a corresponding minimum density that varies only
weakly with true cloud column density or mass, or with ambient pressure in the case of
self-gravitating clouds. Then the characteristic free fall time is also nearly
invariant for that region and the star formation rate per unit dense gas will be about
constant. Such near-constancy will be observed even if the actual free fall rate scales
with the square root of local density.

The appearance of a threshold density for star formation was attributed to the likely
association between new-born stars and nearby gas with a free fall time comparable to
the star's age. The association between young stars and high densities gets even
stronger as the ambient density increases or the ages of the observable stars
decreases. There is no physical threshold for star formation in this model, only an
apparent one resulting from the timing of collapse. When combined with the nearly
invariant collapse time in gas selected by density or column density, the observation
of a universal star formation rate per unit dense gas results. Such a universal
specific rate has no physical basis, however, and is even contrary to the observation
that stars form in self-gravitating clouds, considering that the degree of self-gravity
follows from a balance between column density and ambient pressure.

We conclude that star formation can be a continuous and dynamical process, with no
activity thresholds in density or column density, and no special physics or universal
rates in dense gas. The observation of star formation is fraught with selection
effects, however, and this gives the KS relation many different forms.

The dynamical model gives some insight into what might be different in ultra diffuse
galaxies \citep{vandokkum15}, damped Lyman-$\alpha$ galaxies \citep{rafelski16}, and
other seemingly failed systems where star formation rates are extremely low for the
amount of gas present \citep[e.g.,][]{filho16,lisenfeld16}. If the basic ingredient for
the dynamical model is supersonic turbulence generated by pervasive gravitational
instabilities, then either these systems are stable, perhaps because they are thick or
relatively fast-rotating, or their gas motions are subsonic, perhaps because there is
no cool neutral phase. Because metallicity effects do not seem to matter for the
cessation of star formation \citep{rafelski16}, or for the star-formation rates in
general in the KS relation (Section \ref{dirrsf}), H$_2$ formation is apparently not
involved, nor is it a necessary precursor to star formation. Lack of a cool neutral
phase occurs when the pressure is very low for the ambient radiation field
\citep{elmegreen94,wolfire03,schaye04,kanekar11,krumholz13}. This would seem to be a
natural state for galaxies at very low surface densities, i.e., low pressures, where
there is faint background radiation from cosmological sources and an early generation
of stars.

I am grateful to Dr. Ralf Klessen for discussions at an early stage of this research
and to the referee for a careful reading of the manuscript.

\end{document}